\documentclass[fleqn,usenatbib]{mnras}

\usepackage[T1]{fontenc}
\usepackage{graphicx}%
\usepackage{amsmath,amssymb}%
\usepackage[title]{appendix}%
\usepackage{xcolor}%
\usepackage{longtable}
\usepackage{float}
\title[Chromatic Afterglow of GRB 200829A]{Chromatic Afterglow of GRB 200829A}

\author[N. S. Pankov et al.]{
N. S. Pankov,$^{1,2}$\thanks{E-mail: npankov@hse.ru}
A. S. Pozanenko,$^{2}$
P. Yu. Minaev,$^{2,3}$
S. O. Belkin,$^{1,2}$
A. A. Volnova,$^{2}$
I. V. Reva,$^{4}$
\newauthor
A. V. Serebryanskiy,$^{4}$
M. A. Krugov,$^{4}$
S. A. Naroenkov,$^{5}$
A. O. Novichonok,$^{6}$
A. A. Zhornichenko,$^{6}$
\newauthor
V. V. Rumyantsev,$^{7}$
K. A. Antonyuk,$^{7}$
Sh. A. Egamberdiev,$^{8}$
O. A. Burkhonov,$^{8}$
E. V. Klunko,$^{9}$
\newauthor
A. S. Moskvitin,$^{10}$
I. E. Molotov,$^{11}$
R. Ya. Inasaridze$^{12}$
\\
$^{1}$National Research Institute "Higher School of Economics", ul. Myasnitskaya 21/4, build. 5, Moscow 101000, Russia\\
$^{2}$Space Research Institute, Russian Academy of Sciences, ul. Profsoyuznaya  84/32, Moscow 117997, Russia \\
$^{3}$Lebedev Physical Institute, Russian Academy of Sciences, Leninskiy Prospekt 53, Moscow 119991, Russia\\
$^{4}$Fesenkov Astrophysical Institute, Observatory 23, Almaty 05002, Kazakhstan\\
$^{5}$Institute of Astronomy, ul. Pyatnitskaya 48, Moscow 119017, Russia\\
$^{6}$Petrozavodsk State University, Prospekt Lenina 33, Petrozavodsk 185035, Republic of Karelia, Russia\\
$^{7}$Crimean Astrophysical Observatory, p. Nauchny 185035, Republic of Crimea, Russia\\
$^{8}$Mirzo Ulug'bek nomidagi Astronomiya instituti, Astronomicheskaya str. 33, Toshkent 100052, O`zbekiston\\
$^{9}$Institute of Solar-Terrestrial Physics, Russian Academy of Sciences, Siberian Branch, ul. Lermontova 126a, Irkutsk 664033, Russia\\
$^{10}$Special Astrophysical Observatory, Russian Academy of Sciences, Nizhnii Arkhyz 369167, Karachai-Cherkessian Republic, Russia\\
$^{11}$Keldysh Institute of Applied Mathematics, Miusskaya Ploshchad' 4, Moscow 125047, Russia\\
$^{12}$Evgeni Kharadze Georgian National Astrophysical Observatory, Abastumani 0301, Georgia
}

\pubyear{2015}

\begin{document}
\label{firstpage}
\pagerange{\pageref{firstpage}--\pageref{lastpage}}
\maketitle

\begin{abstract}
    We present the results of our analysis of multiwavelength observations for the long gamma-ray burst GRB 200829A. The burst redshift $z \approx 1.29 \pm 0.04$ has been determined photometrically at the afterglow phase. In gamma rays the event is one of the brightest (in isotropic equivalent), $E_{iso} \gtrsim 10^{54}$ erg. The multicolor light curve of the GRB 200829A afterglow is characterized by chromatic behavior and the presence of a plateau gradually transitioning into a power-law decay that can also be interpreted as a quasi-synchronous inhomogeneity (flare). We assume that the presence of a chromatic inhomogeneity in the early afterglow is consistent with the model of a structured jet.
\end{abstract}


\section{Introduction}\label{sec1}
The cosmological nature of gamma-ray bursts (GRBs) proposed at the beginning of their studies \citep{pri1975,pac1986} has been confirmed by measuring their redshifts. The median redshift for long bursts is $z_{med} = 1.67$ (see, e.g., \citealt{dai2006,tsv2017,tsv2021}). At the same time, the fraction of events with redshifts directly measured through optical spectroscopic observations is $\sim 5\%$ of events \citep{kan2006,but2007,lie2016,kie2020}. There are other known z estimation methods, for example, modeling the photometric afterglow magnitudes \citep{sha2010}, the spectroscopic or photometric method of modeling broadband host galaxy spectra (see, e.g., \citealt{vol2014}).

Long GRBs are associated with the core collapse of massive stars \citep{col1968,bis1975,woo1993,pac1998} and, as a rule, a supernova (SN) is detected in $\sim 10\%$ of the events with a registered optical component, for example, SN 1998bw, SN 2003dh, SN 2013dx, and GRB 181201A \citep{wan1998,woo1999,maz2003,vol2017,bel2020}. However, it is difficult to detect an SN for GRBs at $z \gtrsim 1$, since it becomes too faint. The farthest SN from GRBs was associated with GRB 000911 at $z = 1.06$ \citep{laz2001,mas2005}.

The color evolution between optical and X-ray afterglow light curves detected, for example, for GRB 050525A \citep{oat2011,res2012} and GRB 130831A \citep{pas2016} is not explained by the standard afterglow model \citep{sar1999,pir2004,kum2015}. The afterglow light curve in the standard model decays monotonically (as a power law with an exponent close to unity) and synchronously in the optical and X-ray bands. The chromaticity is explained, for example, using the model of a structured jet, in which the optical and X-ray emissions are simultaneously observed from different jet regions with different Lorentz factors \citep{ben2020,lamb2021,duq2022}. The existence of a structured jet was shown to be possible in numerical simulations (see,
e.g., \citealt{kom2009}).

The X-ray flare can closely follow the flare in the GRB prompt phase, but with a lower amplitude and energy \citep{duq2022}. The presence of a plateau on the light curve is also associated with a small initial Lorentz factor of the jet ($<\Gamma_0> \sim 51$, see \citealt{der2022}) or the formation of a magnetar with a magnetic field $B \sim 10^{15}-10^{16}$ G \citep{pas2007,met2011,row2013}. A correlation between the X-ray plateau duration and the burst luminosity was found by \cite{dai2008,dai2021,dai2022}.
The evolution of the views about GRBs and relevant reviews can be found, for example, in \cite{roz1983,luc1996,pos1999,lev2018,poz2021,yu2022}.

An extensive study of GRBs in the gamma-ray, X-ray, and optical bands was initiated by the Swift observatory \citep{hill2006}. GRB 200829A is one of the brightest GRBs detected by this observatory.
The multiwavelength light curve of the early afterglow from this burst has two inhomogeneities: a plateau and a chromatic flare. The plateau can be associated with the formation of an unstable magnetar with a strong magnetic field. The chromatic behavior of the flare probably arises when observing different jet zones in the X-ray and optical bands, closer to or farther from the jet axis, respectively.
The paper is organized as follows. First, we present the results of the observations with space observatories in which the GRB was detected and the first ground-based observations of its optical
afterglow. Subsequently, we present the results of processing the data obtained with space gamma-ray telescopes in the burst prompt phase. GRB 200829A is one of the most powerful (in isotropic equivalent) GRBs in the energy range 1 keV–10 MeV. In the next section we provide the observational data for the afterglow in the optical and X-ray bands and construct its light curves. Their joint analysis showed the presence of a plateau and a chromatic flare in
the early afterglow. Then, we present the procedure and the redshift estimate obtained by modeling the multicolor light curve of the early afterglow in the optical and X-ray bands. The redshift estimate for
GRB 200829A is $z = 1.29 \pm 0.04$. The results of the detection and analysis of the observations of the host galaxy follow next. We obtain estimates of the absorption parameters, the star formation rate, and the mass of the GRB 200829A host galaxy consistent with the parameters of other host galaxies of long GRBs. In the next section we present the results of our analysis of the chromatic inhomogeneity
and the plateau on the light curve. Our estimates of the physical parameters are consistent with the model of a structured jet \citep{ben2020,lamb2021,duq2022} as a cause of the chromatic behavior. The plateau end time and the isotropic energy in the 0.3–10 keV band are in agreement with the correlation presented by \cite{dai2008,dai2021,dai2022}. In the final section we systematize the results of the observations and their analysis.

The luminosity distance $D_L$ in this paper was calculated using the $\Lambda$CDM model of the Universe with the following cosmological parameters: $H_0 = 69.6 \pm 0.7$ km s$^{-1}$ Mpc$^{-1}$, $\Omega_m = 0.286 \pm 0.008$, and $\Omega_\Lambda = 0.714 \pm 0.008$ \citep{benn2014}. The statistical errors are given at the $1\sigma$ (66.7\%) confidence level, unless stated otherwise.

\section{Observations}\label{sec:observations}

\subsection{Detection and the first observations}\label{subsec:obs_first_det}

GRB 200829A was detected as a bright GRB of duration $\sim 30$ s with the BAT instrument onboard the Swift space observatory; the report on its detection was distributed via the GCN/TAN\footnote{\url{https://gcn.nasa.gov}} automated alert system \citep{sie2020}. The Swift observatory also detected an X-ray component of GRB 200829A with the XRT (0.3–10 keV). The X-ray source was found on August 29, 2020, at 14:01:43.1 UT, i.e., 128.7 s after the BAT trigger at $27^{\prime\prime}$ from the center of the error circle for the gamma-ray component \citep{sie2020,goa2020}.

The Swift Ultraviolet/Optical Telescope (UVOT) began to observe the GRB 200829A field almost immediately after the XRT. It detected a bright optical afterglow with a magnitude $white = 14.28 \pm 0.14$ \citep{sie2020} at coordinates (J2000) R.A. = 16:44:49.14 and Dec. = +72:19:45.63 with a statistical error of $\pm 0.35^{\prime\prime}$ (90\% confidence; \citealt{kui2020}). In the subsequent UVOT observations the source continued to fade \citep{kui2020}.

\subsection{Ground observations}\label{subsec:obs_ground}

\begin{figure}
\centering
\includegraphics[width=0.45\textwidth]{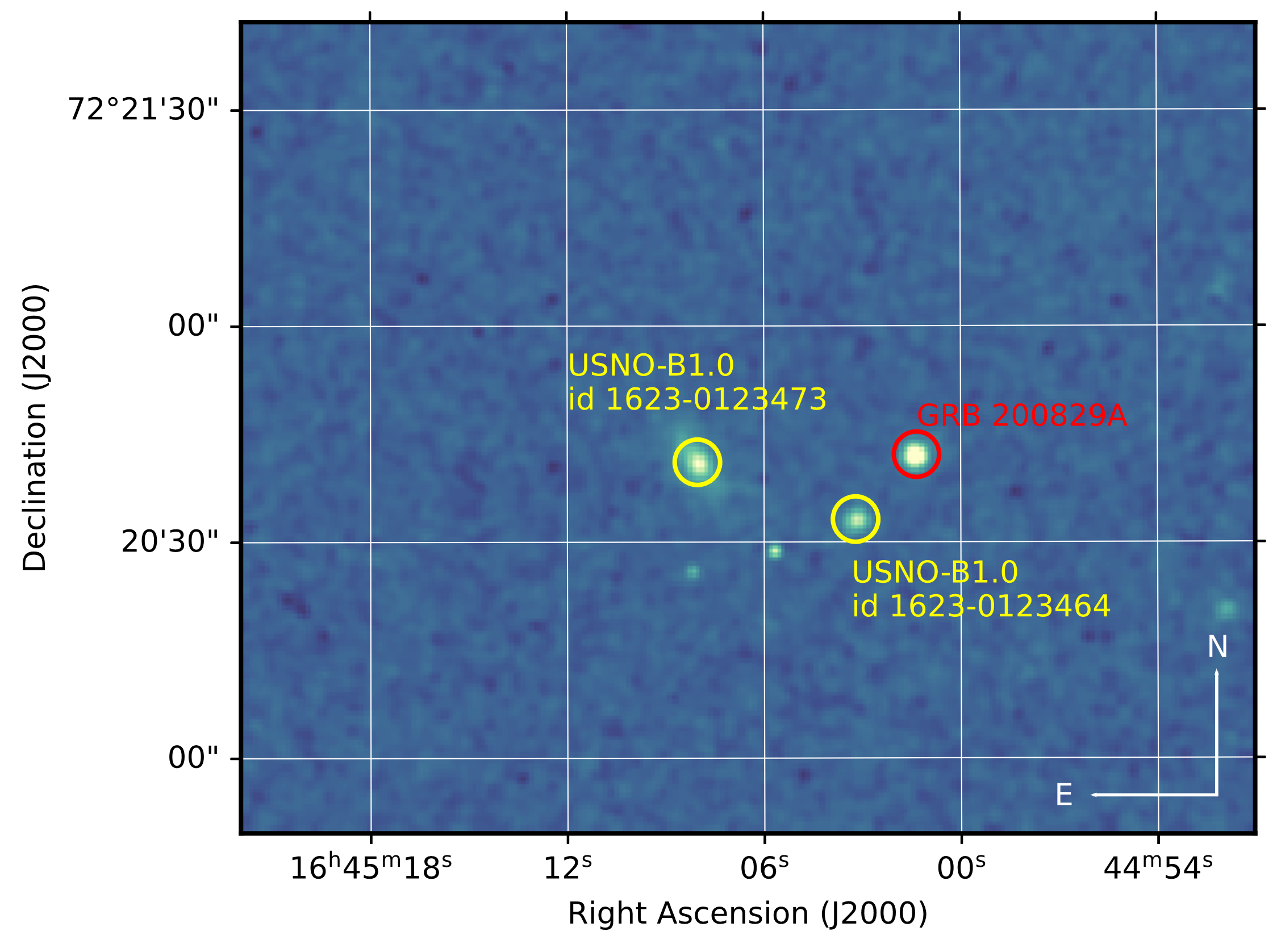}
\caption{Image of the GRB 200829A (highlighted by the red circle) field taken with TShAO Zeiss-1000 on August 29, 2020, at 14:49:05 UT. The catalogued USNO-B1.0 objects are marked by the yellow circles and are labeled in the figure. In the image the north is upward and the east is leftward.}\label{fig:grb_findchart}
\end{figure}

The ground-based optical observations of the fading afterglow of GRB 200829A began $\sim$ 1 hr after the BAT trigger \citep{poz2020} with the 1-meter Zeiss-1000 telescope of the Tien-Shan Astronomical Observatory (TShAO) as the part of GRB-IKI-FuN network (Volnova et al. 2021). One hour after the trigger the magnitude of the source was $R = 16.8 \pm 0.1$ in a single image (Fig.\ref{fig:grb_findchart}) with an exposure time of 60 s.

The observations for 2.83 h, beginning from September 29, 2022, 14:49:05 UT, with the Zeiss-1000 telescope at TShAO allowed a detailed R-band light curve with a time resolution $\sim 60$ s to be constructed. Other instruments of the GRB-IKI-FuN network were additionally used to observe the fading afterglow. For example, the observations with the RC-36 telescope of the Kitab Observatory were
carried out without a filter quasi-synchronously with TShAO. Beginning from August 29, 2020, 17:25 UT, telescopes on the Crimean Peninsula joined the observations: Zeiss-1000 at the observatory on Mount Koshka in the R and I filters, AZT-11 at the Crimean Astrophysical Observatory (CrAO) in the R filter. Owing to the coordinated work of several telescopes, a detailed light curve was constructed over 12 h of observations. One day after the trigger the optical transient was observed with the AS-32 telescope of the Abastumani Astrophysical Observatory (AbAO) in the R filter and the Zeiss-1000 telescopes (Mount Koshka and the Special Astrophysical Observatory of the Russian Academy of Sciences (SAO RAS) in the R and I filters). Upper limits in the R filter were also obtained on the third and fifth days after the trigger in the observations with the AZT-33IK telescope at the Mondy observatory of the Institute of
Solar-Terrestrial Physics, the Siberian Branch of the Russian Academy of Sciences. When the afterglow could no longer be observed with other telescopes due to the source brightness decline, $R \gtrsim 22^{m}.5$, the AZT-22 telescope of the Maidanak Astronomical Observatory (MAO) was used. MAO is located in a unique climatic region and, for this reason, the seeing (angular resolution) here reaches $0^{\prime\prime}.7$. A total of four deep upper limits were obtained in the $R$ filter: three in the period from 13 to 45 days after the burst.

\section{Prompt emission phase (gamma-rays)}\label{sec:prompt}

\subsection{GBM/Fermi data analysis}\label{subsec:prompt_gbm}
In our analysis of GRB 200829A we used the publicly accessible data of the GBM/Fermi experiment\footnote{\url{ftp://legacy.gsfc.nasa.gov/fermi/data/}}.

\subsubsection{Light curve}
\begin{figure}
\centering
\includegraphics[width=0.45\textwidth]{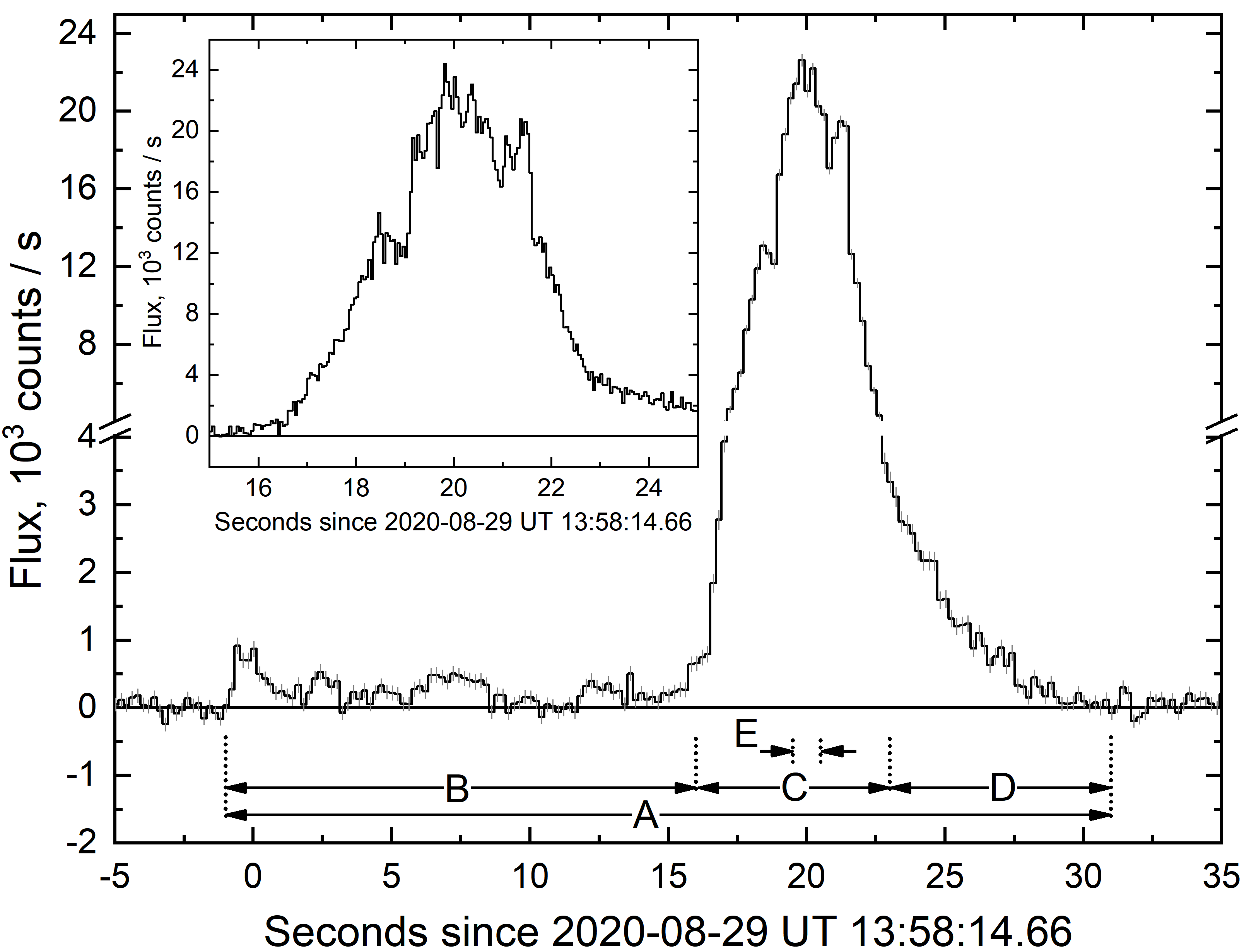}
\caption{Light curve of GRB 200829A in the energy range 7–850 keV with a time resolution of 0.5 s (from GBM/Fermi data). The observed flux in counts per second over the background model is along the vertical axis; the time since the GBM/Fermi
trigger in seconds is along the horizontal axis. The $1\sigma$ flux errors are shown. The vertical dotted lines mark the boundaries of the intervals used in our spectral analysis (see Table 1). The inset shows the light curve of the main episode with a time resolution of 0.05 s.}\label{fig:gbm_v2}
\end{figure}

The light curve in the energy range 7-850 keV constructed using data from the NaI04 and NaI08 detectors of the GBM/Fermi experiment most illuminated by the GRB 200829A source is presented in Fig.\ref{fig:gbm_v2}. 

The light curve can be arbitrarily divided into two activity episodes of approximately equal duration: the dim initial and main bright ones. The total duration of the gamma-ray emission from GRB 200829A in the GBM/Fermi experiment is more than 30 s, with the formal duration parameters $T_{90} = 8.4 \pm 0.1$ s and $T_{50} = 2.5 \pm$ 0.1 s \citep{kosh96} covering only the brightest part of the main episode. The values obtained characterize GRB 200829A as a long burst (type II, a collapsar) (see, e.g., \citealt{kou93,min10b,min17}). 

As can be seen from Fig.\ref{fig:gbm_v2}, both initial and main burst episodes have a complex structure that consists of a large number of overlapping pulses whose parameters, therefore, are impossible to reveal. Due to the superposition of pulses, analyzing the spectral evolution by the method of cross-correlation analysis for this burst is meaningless (see, e.g., \citealt{min14}).

\subsubsection{Spectral analysis}

To construct and fit the energy spectra, we used the RMfit v4.3.2 software package specially developed to analyze the GBM/Fermi data\footnote{\url{http://fermi.gsfc.nasa.gov/ssc/data/analysis/rmfit/}}. The technique of spectral analysis, including the choice of an optimal spectral model, is similar to that used in \cite{grub14}. We analyzed the energy spectra based on data from the most illuminated
$NaI_{04}$, $NaI_{08}$, $BGO_{00}$, and $BGO_{01}$ detectors.

We analyzed the energy spectrum of
GRB 200829A in five different time intervals (Fig.\ref{fig:gbm_v2}). Intervals A, B, C, D, and E correspond to the integrated spectrum of the event, the initial dim episode, the main part of the main episode, the decay stage of the main episode, and the peak in the light curve (the peak flux on a time scale of 1 s), respectively. 

The energy spectrum of all the investigated components of GRB 200829A is unsatisfactorily described by both simple power-law model and thermal model; the optimal model is a power law with an exponential cutoff (CPL) or a broken power law (\citealt{band93}). The results of our spectral analysis (the parameters of the optimal spectral models) are presented in Table \ref{tab:gbmspectra}.

\begin{table*}[t]
\centering
\caption{Results of our spectral analysis of GRB 200829A in the gamma-ray band based on GBM/Fermi data}
\tiny

\begin{tabular}{lccccccc} 
\\
\hline
Interval $^a$ & Spectral & $\alpha$ & $\beta$ & $E_p$$^c$ & Fluence $^d$ & E$_{iso}$ $^e$ & $EH$ $^{f}$ \\
(c) & model $^b$ & & &(keV) & ($10^{-5}$ erg cm$^{-2}$) & ($10^{53}$ erg) & \\ \hline
%\startdata

A = -1 -- 31 & Band & $-0.552\pm0.014$ & $-2.468\pm0.026$  & $357\pm5$ & $22.17\pm0.08$ & $13.02\pm0.13$ & 0.45 \\ 
B = -1 -- 16 & CPL & $-1.23\pm0.10$ & --  & $249\pm37$ & $0.68\pm0.04$ & $0.323\pm0.025$ & 1.42 \\ 
C = 16 -- 23 & Band & $-0.392\pm0.013$ & $-2.477\pm0.016$ & $363\pm4$ & $20.55\pm0.06$ & $12.16\pm0.08$ & 0.47 \\ 
D = 23 -- 31 & Band & $-0.89\pm0.10$ & $-2.30\pm0.09$  & $136\pm13$ & $1.189\pm0.026$ & $0.66\pm0.04$ & 0.57 \\ 
E = 19.5 -- 20.5 & Band & $-0.333\pm0.023$ & $-2.288\pm0.020$ & $393\pm8$ & $6.011\pm0.033$ & $3.88\pm0.04$ & 0.79 \\ 
\hline

\multicolumn{8}{l}{$^{a}$ The time interval relative to the GBM/Fermi trigger.}\\
\multicolumn{8}{l}{$^{b}$ CPL is a power law with an exponential cutoff, Band is a broken power law (Band et al. 1993).}\\
\multicolumn{8}{l}{$^{c}$ The peak energy in the energy spectrum $\nu F \nu$.}\\
\multicolumn{8}{l}{$^{d}$ The fluence in the range 10 -- 1000 keV in the observer frame.}\\
\multicolumn{8}{l}{$^{e}$ The isotropic equivalent energy in the range 1 keV - 10 MeV in the source frame ($z = 1.29 \pm 0.04$, see below).}\\
\multicolumn{8}{l}{$^{f}$ The classification parameter $EH$ (Eq~\ref{eq:eh}).}\\

%\enddata
\label{tab:gbmspectra}
\end{tabular}
\end{table*}

\begin{figure}
\centering
\includegraphics[width=0.45\textwidth]{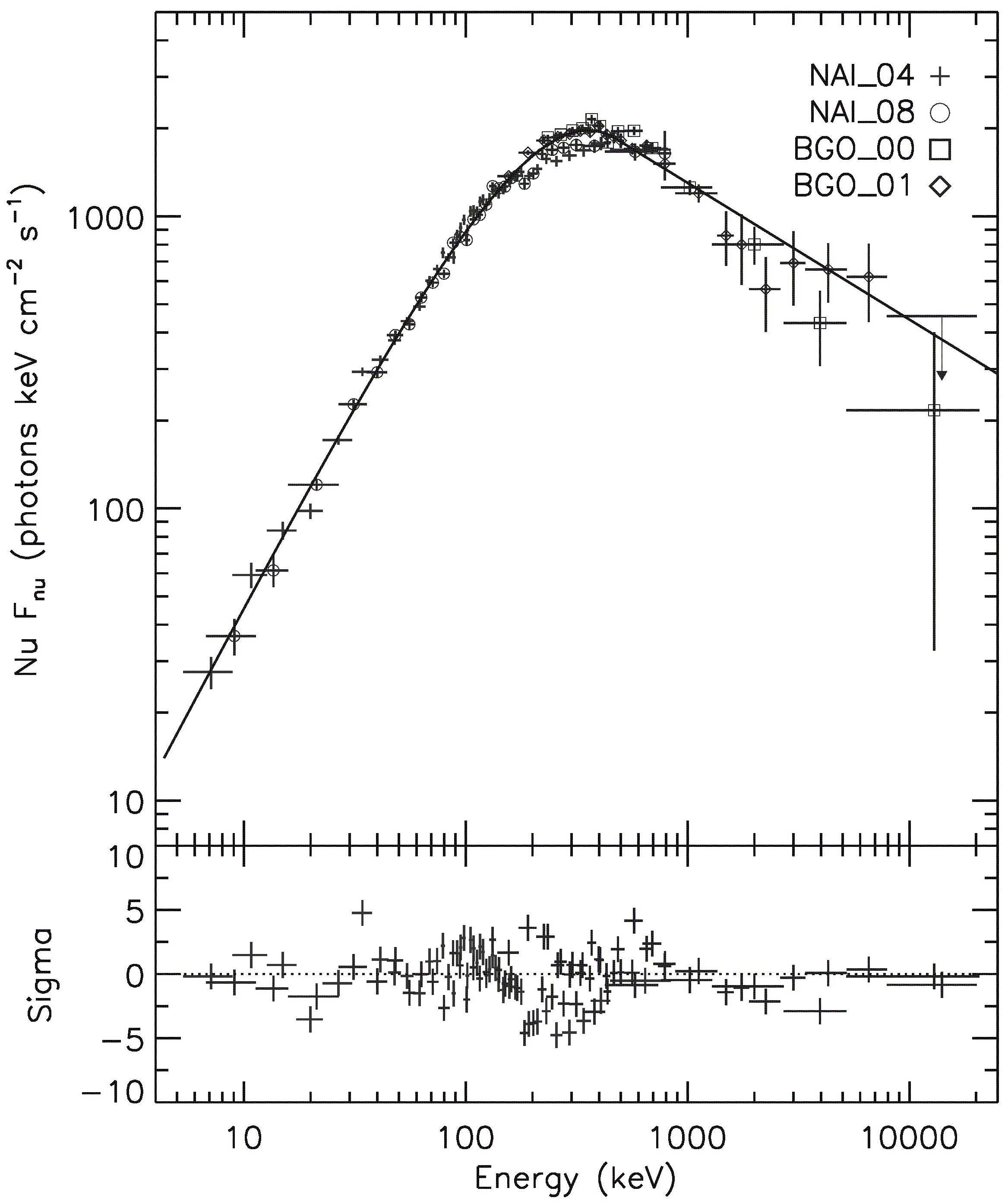}
\caption{The energy spectrum $\nu F \nu$ of GRB 200829A constructed using data from the $NaI_{04}$, $NaI_{08}$, $BGO_{00}$, and $BGO_{01}$ detectors of the GBM/Fermi experiment in the time interval A = –1 -- 31 s relative to the trigger covering the GRB prompt phase (the upper part of the figure). The smooth curve indicates a broken power-law spectral fit \citep{band93}). The deviation of the spectral model from the experimental data expressed in standard deviations is shown in the lower part.}\label{fig:gbm_spec}
\end{figure}

As an example, Fig.\ref{fig:gbm_spec}  shows the integrated energy spectrum fitted by a broken power law (\citealt{band93}) with the parameter $E_p = 357 \pm 5$ keV characterizing the position of the peak in the energy spectrum $\nu F_{\nu}$ . The derived parameters of the energy spectrum for all the investigated components of the light curve are typical for GRBs (see, e.g., \citealt{grub14}). It can be seen from Fig. 3 that the scatter of data points relative to the model is fairly large.
This is apparently due to the systematic effects (for example, inaccurate response matrices and detector cross-calibration) that arise when reconstructing the photon spectrum with the RMfit v4.3.2 software package.

\subsubsection{$E_{p,i}$ -- $E_{iso}$ correlation}

GRBs are characterized by a number of correlations between various observed parameters. One of the best-known ones is the correlation of the isotropic equivalent energy released in the energy range 1 keV -- 10 MeV, $E_{iso}$ , with the position of the extremum in the energy spectrum $\nu F_{\nu}$ in the source frame, $E_{p,i}$ (\citealt{ama02}). The nature of the correlation is still the subject of discussion. One possible explanation implies viewing angle effects: the smaller the angle between the source–observer line and the jet axis, the brighter and spectrally harder the GRB (\citealt{eic04,lev05,poz18a}). The parameter $E_{iso}$ is calculated from Eq. (\ref{eq:eiso}), where $F$ is the fluence (time-integrated flux) in the energy range 1 keV -- 10 MeV in the source frame, $D_L$ is the luminosity distance to the source, and $z$ is its redshift. The corresponding values of the parameter $E_{iso}$ for GRB 200829A were calculated using
the redshift $z = 1.29 \pm 0.04$ and are given in Table \ref{tab:gbmspectra}. To estimate the fluence in the energy range 1 keV -- 10 MeV, we used an extrapolation of the spectral model to low energies (below 6 keV in the observer
frame), 

\begin{equation}\label{eq:eiso}
    E_{iso} = \frac{4\pi D_L^{~2}F}{1+z}.
\end{equation}

\begin{figure}
\centering
\includegraphics[width=0.45\textwidth]{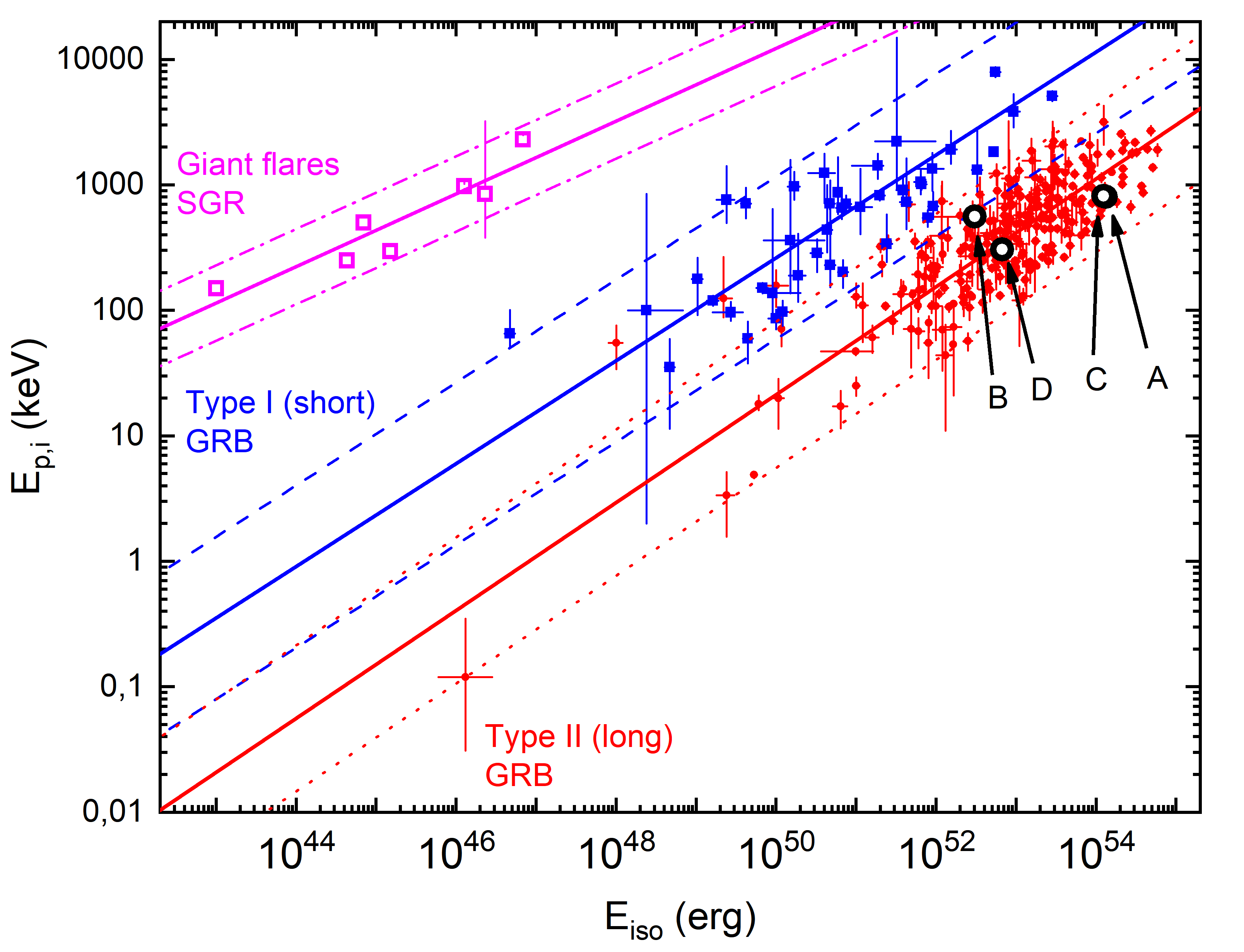}
\caption{$E_{p,i}$ -- $E_{iso}$ diagram for type I (blue squares) and type II (red circles) GRBs and SGR giant flares (magenta unfilled squares) with the corresponding fits (solid lines) and $2 \sigma$ correlation regions (dashed lines). The black unfilled circles indicate the positions of GRB 200829A and its individual episodes (see Table \ref{tab:gbmspectra}).}\label{fig:amati}
\end{figure}

Fig.\ref{fig:amati} presents a $E_{p,i}$ -- $E_{iso}$ diagram for one of the most complete samples of 317 GRBs with known redshifts and determined parameters $E_{p,i}$ published in \cite{min20b,min21} and for 7 giant flares from soft gamma repeaters (SGRs), for which the $E_{p,i}$ -- $E_{iso}$ correlation was first found by \cite{min20a}.

On the $E_{p,i}$ -- $E_{iso}$ diagram GRB 200829A occupies a position typical for long (type II) GRBs, while being one of the brightest bursts with $E_{iso} > 10^{54}$ erg. The individual burst episodes are also located in the correlation region of long GRBs.

\subsubsection{$T_{90,i}$ -- $EH$ diagram}

The $E_{p,i}$ -- $E_{iso}$ correlation can also be used for the classification of GRBs, since the correlation region of type I (short) GRBs is above the correlation region of type II (long) GRBs, with the correlation for both types of GRBs being described by a power law with a single index, $\alpha = -0.4$ \citep{min20a,min20b}. For this purpose, \cite{min20b} introduced the parameter EH (Eq. \ref{eq:eh}) that characterizes the GRB position on the $E_{p,i}$ -- $E_{iso}$ diagram. Compared to type II GRBs, type I GRBs have a greater spectral hardness $E_{p,i}$ at a lower total energy $E_{iso}$ and, as a consequence, a larger value of the parameter $EH$:

\begin{equation}
    EH = \frac{(E_{p,i} / 100~{keV})}{ (E_{iso} / 10^{51}~{erg})^{~0.4}}.
	\label{eq:eh}
\end{equation}

\begin{figure}
\centering
\includegraphics[width=0.45\textwidth]{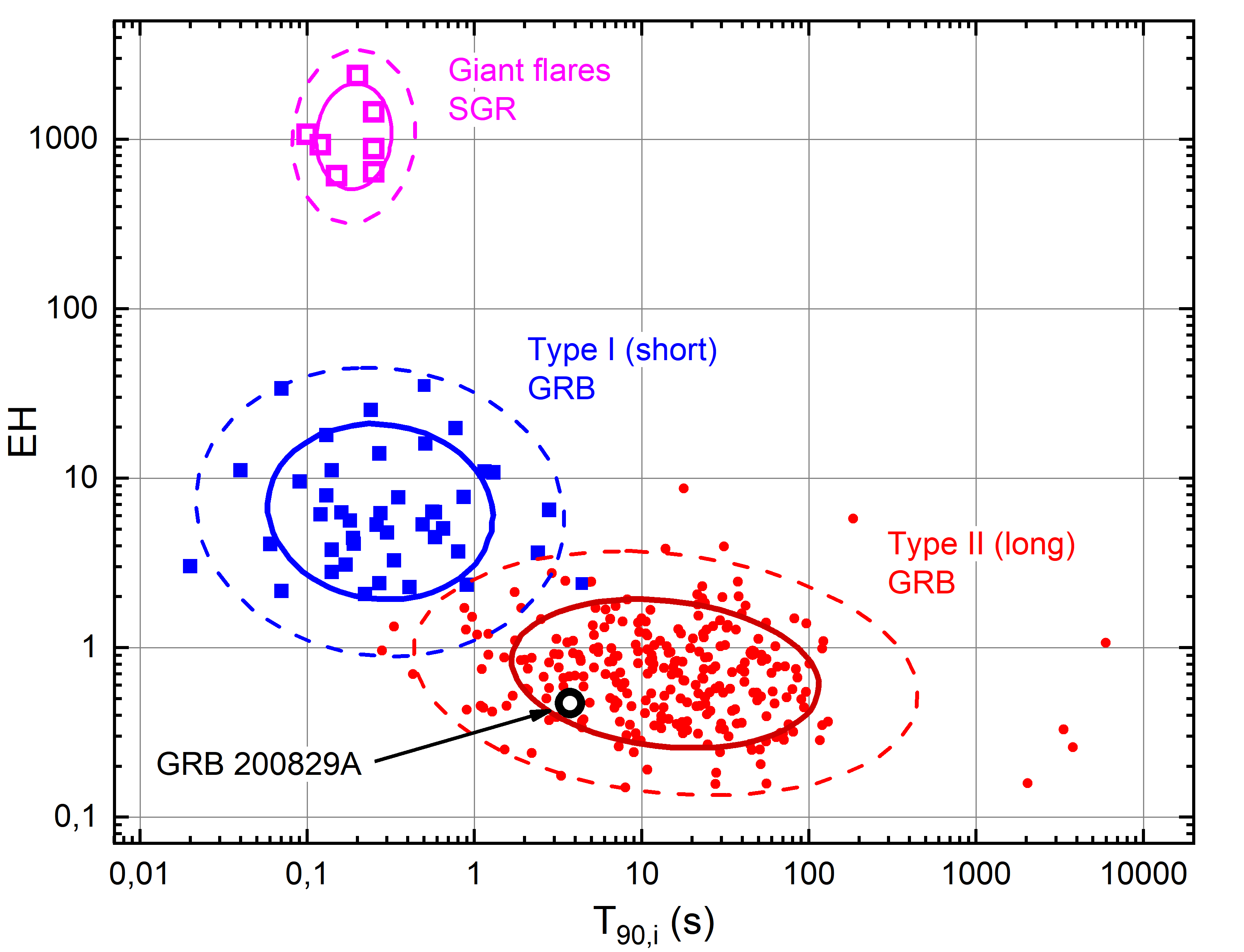}
\caption{$T_{90,i}$ -- $EH$ diagram for type I (blue squares), type II (red circles) GRBs, and giant SGR flares (pink open squares) with the corresponding results of a cluster analysis (the $1\sigma$ and $2\sigma$ cluster regions are indicated by the thick solid and thin dashed curves, respectively). The black unfilled circle indicates the position of GRB 200829A.}\label{fig:ehd}
\end{figure}

The most efficient GRB classification method suggests a joint analysis of the parameter $EH$ and the duration parameter $T_{90,i}$ measured in the source frame \citep{min20a,min20b,min21}. Fig.\ref{fig:ehd} presents a $T_{90,i}$ -- $EH$ diagram for 317 GRBs and 7 giant SGR flares from \cite{min20a,min20b,min21} that provides the best
separation into clusters of the corresponding types of transients (the smallest overlap region) among the known classification schemes.

GRB 200829A being investigated is located within the $1\sigma$ cluster region of type II (long) GRBs, although
it has a relatively small value of the duration parameter, $T_{90,i} = 3.7$ s, corresponding to the overlap region of the duration distributions for the classes of long and short bursts, emphasizing the stability of the classification system based on the $T_{90,i}$ -- $EH$ diagram method. The values of the parameter $EH$ for GRB 200829A and its individual episodes are given in Table \ref{tab:gbmspectra}.

\subsection{BAT/Swift Data Analysis}

As has been shown above, GRB 200829A is characterized by a fairly hard energy spectrum with $E_p > 350$ keV located outside the effective BAT/Swift sensitivity range, 15 -- 150 keV. Therefore, in this section devoted to analyzing the burst based on BAT/Swift data, we restricted our study only to its light curve. 

As the data source we used the publicly accessible service\footnote{\url{https://www.swift.ac.uk/burst_analyser/}}. Although the trigger time in the BAT/Swift experiment differs by 79.74 s from that in GBM/Fermi, as the trigger time $T_0$ we will use the trigger time of the latter, since it corresponds more closely to the start time of the burst prompt phase (Fig.\ref{fig:gbm_v2}).

\begin{figure}
\centering
\includegraphics[width=0.45\textwidth]{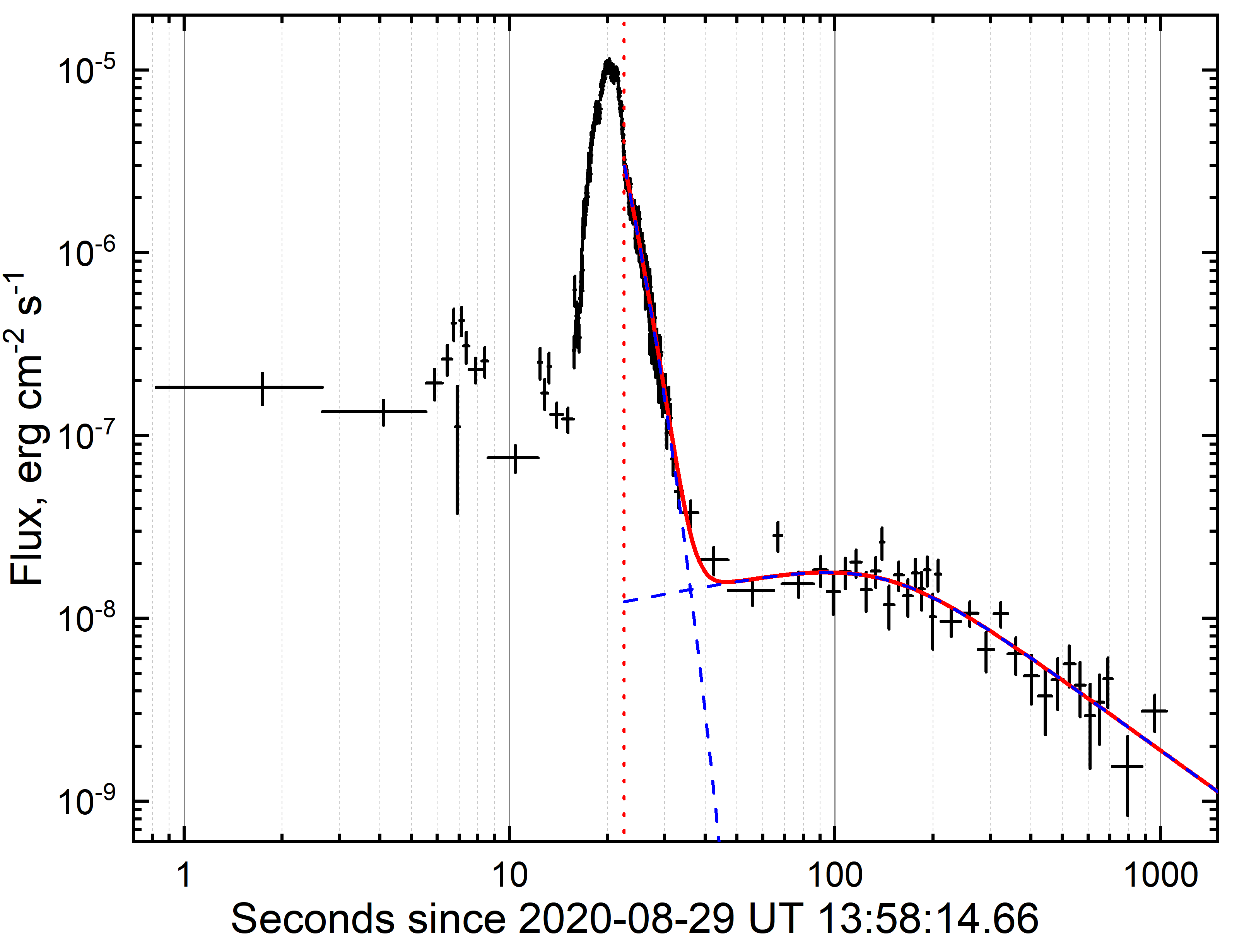}
\caption{Light curve of GRB 200829A in the 15 -- 50 keV energy band (from BAT/Swift data). The energy flux in erg cm$^{-2}$ s$^{-1}$ is along the vertical axis; the time since the GBM/Fermi trigger in seconds is along the horizontal axis. The $1\sigma$ flux errors are shown. The light curve is fitted by the sum of an exponential model and a broken power law (blue dashed lines). The red solid line is their sum; the vertical red dotted straight line marks the beginning of the time interval in which the fit was constructed.}\label{fig:bat}
\end{figure}

The light curve constructed in the 15 -- 50 keV energy band is presented in Fig.\ref{fig:bat}. After binning the light curve by the signal accumulation method until a certain statistical significance was reached, we detected a weak, but statistically significant extended emission with a duration $\sim 1000$ s that is a separate component of the light curve.

The light curves of individual pulses at the prompt phase of GRBs are usually characterized by the so- called FRED shape (fast rise -- exponential decay, see, e.g., \citealt{nor05,hakk11,min14}). The light curve of GRB 200829A has a complex shape and is a superposition of several significantly overlapping FRED pulses, making it difficult to fit the light curve of the entire prompt phase. Therefore, when jointly fitting the component of the
prompt phase and the extended emission, we used only the decay stage of the last pulse of the prompt phase that was described by an exponential model.

The light curve of the extended emission is characterized by an initial quasi-plateau stage followed by a power-law flux decline. To describe this component of the light curve, we used a broken power law \citep{beu1999}.

The results of jointly fitting the decay stage of the prompt phase and the extended emission component by the above models are presented in Fig.\ref{fig:bat}. At the initial stage the index of the light curve for the extended emission is $\alpha = 0.34 \pm 0.27$, which does not rule out the plateau stage ($\alpha = 0$), the break in the light curve is observed at $t_{br} = 143 \pm 38$ s, and the index after the break is $\beta = -1.28 \pm 0.18$. The index after the break is close to a value typical for
an afterglow ($\beta \sim -1$), suggesting a connection of the extended emission component with the afterglow, which is typical for bright GRBs (see, e.g., \citealt{mozg21}). In our further analysis of the X-ray and optical data this interpretation will be confirmed.

\subsection{SPI-ACS/INTEGRAL Data Analysis}

GRB 200829A was also detected by the anticoincidence shield (ACS) of the SPI gamma-ray spectrometer onboard the INTEGRAL observatory \citep{kien03}.

\begin{figure}
\centering
\includegraphics[width=0.45\textwidth]{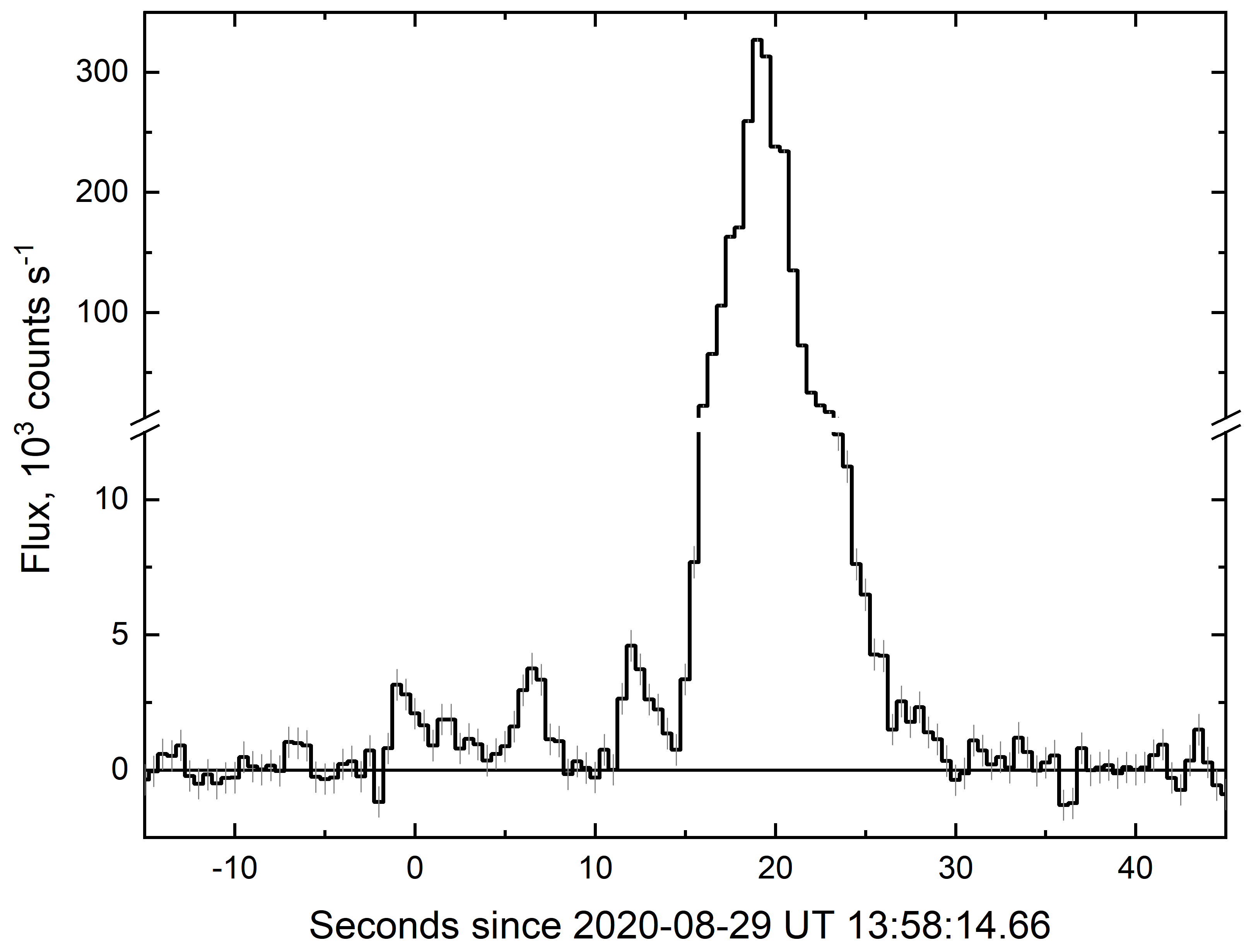}
\caption{Light curve of GRB 200829A at energies above 80 keV with a time resolution of 0.5 s (from SPI-ACS/INTEGRAL data). The observed flux in counts per second over the background model is along the vertical axis; the time since the GBM/Fermi trigger in seconds is along the horizontal axis. The $1 \sigma$ flux errors are shown.}\label{fig:acs}
\end{figure}

The light curve of GRB 200829A at energies above 80 keV with a time resolution of 0.5 s constructed from the SPI-ACS/INTEGRAL data\footnote{\url{http://isdc.unige.ch/\~savchenk/spiacs-online/spiacs-ipnlc.pl}} is presented in Fig. \ref{fig:acs}. Despite the stable back-ground conditions in the time interval from -5000 to 5000 s around the GRB typical for the SPI- ACS detector \citep{min10a,mozg21}, no extended emission was detected by SPI-ACS/INTEGRAL, although it was recorded in the BAT/Swift experiment with great confidence.

This is probably due to the relatively soft energy spectrum (the small fraction of high-energy emission) of this component and the high lower sensitivity threshold of the ACS detector (80 keV). A possible precursor \citep{min17} was not detected either. Although several dim episodes with a total duration of about 15 s, from which the burst prompt phase begins, were recorded in the SPI-ACS experiment (Fig. \ref{fig:acs}), the duration parameters $T_{90}$ and $T_{50}$ cover only the main bright episode and are $T_{90} = 5.5 \pm 0.1$ s and $T_{50} = 2.1 \pm 0.1$ s, confirming the classification of this burst as type II (long).

According to the SPI-ACS/INTEGRAL data, the time-integrated flux from GRB 200829A was $F = (115.3 \pm 0.3) \times 10^4$ counts. Using the cross-calibration of the SPI-ACS and GBM experiments for GRBs from \citealt{poz2020}, we obtain an estimate of the fluence in the energy range 10 -- 1000 keV, $F \gtrsim 2.9 \times 10^{-4}$ erg cm$^{-2}$.

Given the systematic calibration error for the SPI-ACS response to GRB energy spectra differing in shape, the estimated fluence can vary within the range $8.4 \times 10^5$ -- $9.8 \times 10^{-4}$ erg cm$^{-2}$ (at $2 \sigma$ confidence). The estimated fluence agrees well with the value obtained through our spectral analysis of the GBM/Fermi data (Table \ref{tab:gbmspectra}).

\section{Afterglow}

In our analysis we used the public \texttt{Swift Burst Analyzer}\footnote{\url{https://www.swift.ac.uk/burst_analyser/00993768}} data \citep{eva2007,eva2009} from the BAT, XRT, and UVOT experiments onboard Swift to construct the gamma-ray, X-ray, and optical light curves, respectively. We also used the public observational data accessible via the GCN/TAN service\footnote{\url{https://gcn.gsfc.nasa.gov/other/200829A.gcn3}}: the observations with the RC80 robotic telescope at the Konkoly Observatory \citep{vin2020} in the $r^\prime$, and $i^\prime$ filters, the observations with the Zeiss-1000 SAO RAS telescope in the $R_c$ and $I_c$ filters, the observations with the NEXT telescope \citep{zhu2020,zhu2020b} and the observations with the Liverpool telescope \citep{izz2020} in the $g^\prime$, $r^\prime$, $i^\prime$, and $z^\prime$ filters. The publicly available images from the Nordic Optical Telescope (NOT) in the $r^\prime$ filter were retrieved from the \texttt{NOT FITS Header Archive}\footnote{\url{http://www.not.iac.es/observing/forms/fitsarchive/}}. The light curves were fitted using the \texttt{lmfit}\footnote{\url{https://github.com/lmfit/lmfit-py}} package \citep{new2021} for python.

\subsection{Analysis of the optical observations}

The afterglow of GRB 200829A was observed for the first $\sim 3$ days in the X-ray and optical bands. To construct the optical light curve of the GRB 200829A afterglow, we processed the GRB-IKI-FuN observational data and the publicly accessible Nordic Optical Telescope (NOT) data using the developed software pipeline \citep{pan2022} based on the \texttt{APEX} package \citep{kou2012,dev2010}. The standard data processing procedure includes the calibration and quality control of the original images, their alignment and stacking (if required), the extraction of objects in the images, the astrometry and photometry, the construction of a local catalog, the search for and identification of transients in it. In addition to the software pipeline \citep{pan2020}, we used the \texttt{PyRAF} software package \citep{stsci2012} in individual cases, for example, when an optical transient had a low signal-to-noise ratio, $S/N < 5$. We performed the astrometry of images based on the USNO-B1.0 reference catalog \citep{mon2003}. The photometric calibration stars were chosen by cross-matching the stars from USNO-B1.0 with the PanSTARRS PS1 catalog \citep{cha2016}. This allowed us to calibrate the source magnitudes in the images
taken both in $R$, $I$, and in $r$ using the same stars. The photometry of images without a filter was performed relative to the R magnitude. A log of optical observations is given in Appendix A. Figure \ref{fig:multicolor_lc} presents a multicolor optical light curve; for clarity, the magnitudes in all filters, except $R$, were separated relative to their initial values.

\begin{figure}
\centering
\includegraphics[width=0.5\textwidth]{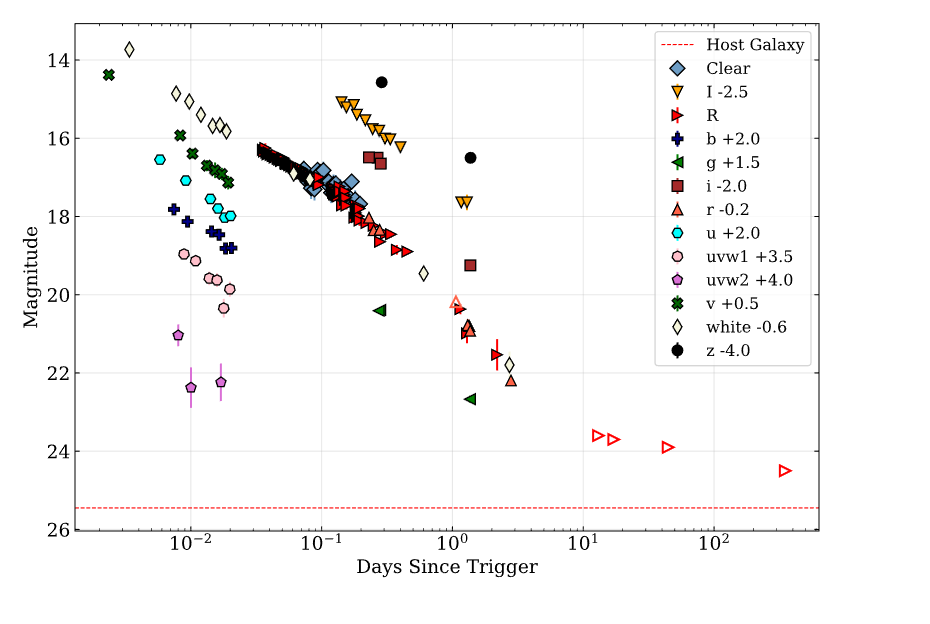}
\caption{Multicolor optical light curve of GRB 200829A. The time since the GBM trigger (days) is along the horizontal axis, and the apparent magnitude is along the vertical axis. The symbols designate the photometric magnitudes obtained in various filters. The dashed line indicates the brightness level of the GRB 200829A host galaxy. The unfilled symbols designate the upper limits (at $3 \sigma$ confidence level).}\label{fig:multicolor_lc}
\end{figure}

A break in the $R$ filter, which may be related to the geometrical viewing angle effect typical for relativistic jets \citep{sar1999,pir2004}, is traceable in Fig.\ref{fig:multicolor_lc}. The magnitudes in the light curve were not corrected for the Galactic extinction characterized by the color excess $E(B - V) = 0.0364$ \citep{sch1998} toward GRB 200829A. Since it is impossible to establish the spectral evolution at the entire optical afterglow stage due to insufficient quasisynchronous coverage in different filters, we assume its absence. Under this assumption the light curves
in the $g$, $r$, $i$, $z$, $I$, $u$, $b$, $v$, $uvw1$, $uvw2$ filters and in $clear$ and $white$ light were reduced to the
R light curve (the data are presented predominantly in this filter) by additionally multiplying the flux by the appropriate numerical coefficient at which the best agreement with the data from the $\chi^2$ test is achieved. In our subsequent analysis of the afterglow we will consider this monochromatic optical light curve.

\subsection{Analysis of the X-ray observations}

The X-ray afterglow was also observed quasisynchronously with the optical one by the XRT/Swift instrument \citep{bur2005}. The XRT is a Wolter I telescope with a focal length of $EFF = 3500$ mm
and a $FOV = 23.6^\prime \times 23.6^\prime$. An E2V CCD detector with $600 \times 600$ pixels and cooled to $-100$ $^{\circ}$C was mounted at the focal plane. As a result, an angular resolution of $\theta \sim 3^{\prime\prime}$ is achieved. The telescope operates in the 0.3 -- 10 keV energy band in several modes, among
which are Windowed Timing (WT), and Photon Counting (PC). In the WT mode a good time resolution ($\sim$ 1.8 ms) is achieved, but, at the cost of, the angular and spectral resolutions. In the PC mode the signal readout rate from the CCD is reduced to $\sim$ 2.5 s, but other characteristics do not change. The WT and PC modes are used at fluxes $F_X = 1 - 600$, and $F_X < 1$ mCrab, respectively \citep{bur2005}.

\begin{figure}
\centering
\includegraphics[width=0.5\textwidth]{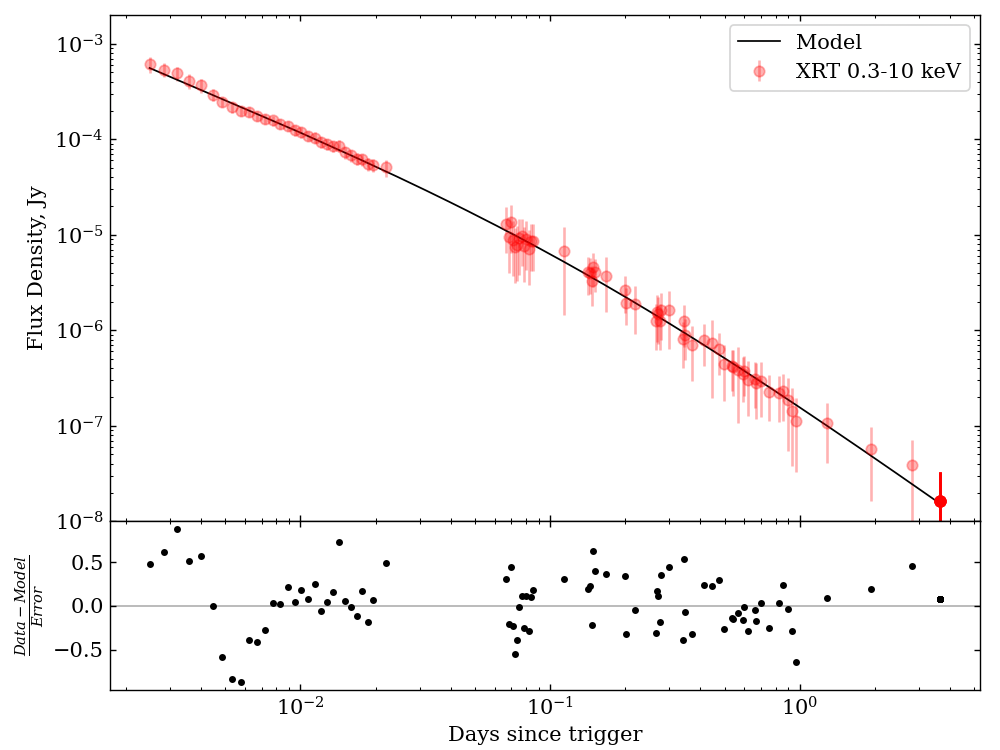}
\caption{The upper panel shows the X-ray light curve of GRB 200829A. The time since the GBM trigger in days is along the
horizontal axis, while the observed flux density in Jy is along the vertical axis. The circles designate the values, while the thick solid line indicates the broken power law fit. The lower panel shows the deviation of the experimental flux densities from the fit in standard deviations $\sigma$.}\label{fig:xray_lc}
\end{figure}

The X-ray light curve of GRB 200829A was constructed from the data publicly available via the \texttt{Swift Burst Analyzer}\footnote{\url{https://www.swift.ac.uk/burst_analyser/00993768/}} service. The data contain the results of observations (time and flux) in the WT and PC modes and span a period $\sim 2$ days. The observed X-ray flux was grouped into bins whose duration increases on a logarithmic scale, starting
from 60 s. The Galactic extinction on the line of sight was not taken into account. Thus, the light curve was smoothed, while the significance of individual measurements was improved. Fig.\ref{fig:xray_lc} presents the X-ray light curve of the GRB 200829A afterglow.

According to Fig. \ref{fig:xray_lc}, a break is traceable in the
X-ray light curve, as in the optical one. Fitting the light curve by the Beuermann function \citep{beu1999} allowed us to determine the slope of the light curve and the position of the break on the
time scale. For example, the slope is $\alpha_1 = 1.06 \pm
0.03$ before the break and $\alpha_2 = 1.88 \pm 0.06$ after the
break. Note that the slope for the X-ray afterglow $\alpha_1 = 1.06 \pm 0.03$ before the break agrees well with the slope for the extended emission in the soft 15 -- 50 keV gamma-ray band $\beta = 1.28 \pm 0.18$ found previously. Hence a correlation of the observed extended emission and the afterglow can be assumed. According to this assumption, the gamma-ray light curve was
rescaled to the X-ray light curve in the time interval 100 -- 1000 s relative to $T_0$ (in which the data from both telescopes were obtained) assuming the afterglow to be achromatic in the X-ray and soft gamma-ray bands. According to the $\chi^2$ test, the conversion coefficient is $k = F_{XRT} /F_{BAT} = 4.54$. Thus, we constructed the most complete light curve of the X-ray
afterglow that will be used in our further analysis. 

\subsection{Multiwavelength light curve}

\begin{figure*}
\centering
\includegraphics[width=0.85\textwidth]{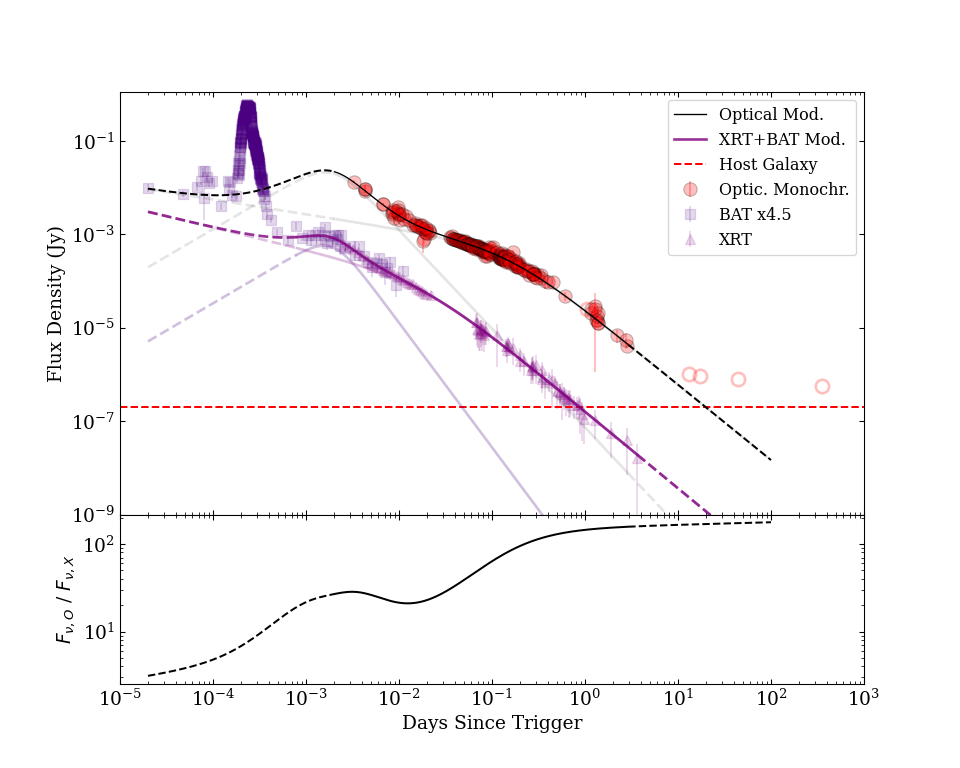}
\caption{The upper panel shows the light curve of GRB 200829A. The time since the GBM trigger $t-T_0$ (days) is along the horizontal axis, while the spectral flux density (Jy) is along the vertical axis. The photometric values in the optical filters (red filled squares) were combined into a single monochromatic curve. The X-ray light curve (violet filled triangles) was combined with the gamma-ray one (dark-blue filled squares) assuming that the extended emission corresponds to the X-ray afterglow. The open symbols designate the upper limits. The model curves are marked by the solid lines. The lower panel shows a graph of the evolution of the optical-to-X-ray flux ratio.}\label{fig:total_lc}
\end{figure*}

Let us investigate the previously constructed X-ray and optical light curves (presented together in Fig. \ref{fig:total_lc}) and fit and interpret them.

In Fig. \ref{fig:total_lc} both X-ray and optical light curves consist of three episodes (not counting the burst prompt phase): a flare gradually decays into a power-law flux decline and then a break after which the flux fades according to a steeper law. At the same time, the leading edge of the flash looks rather steep against the background of the power-law decay. Since in the optical light curve the time of the peak cannot be unambiguously determined due to the absence of data, we assume that it coincides with that for the X-ray band.

To fit the light curves, we chose a function in the form of a sum of two power laws with a smooth break \citep{beu1999} defined by Eq.\ref{eq:ag_model}:

\begin{equation}\label{eq:ag_model}
    F(t) = (F_1^{-n} + F_2^{-n})^{-\frac{1}{n}} + (F_3^{-n} + F_4^{-n})^{-\frac{1}{n}}
\end{equation}

where, $F_i = (\frac{t - T_0}{t_{b_j}})^{\alpha_i}$, $t$ -- the time since trigger (days), $t_{b_i}$ -- the break time (days), $\alpha_i$ -- are the indices before and after the break, $n$ -- the smoothing parameter ($n > 0$), $i = {1, 2, 3, 4}$, $j=1$ at $i=1,2$, and $j=2$ at $i=3,4$. The fitting results are presented in the Table \ref{table:ag_fit}; the values marked by * were fixed. The fitting by the two-component model was performed in the intervals $\sim 50$ s -- 3.66 days in the X-ray band and 283 s -- 2.80 days in the optical band (relative to the GBM/Fermi trigger). Fig.\ref{fig:total_lc} (bottom) demonstrates the color evolution between the X-ray and optical bands.

\begin{table*}
\centering
\caption{Afterglow fitting results}
\begin{tabular}[H]{p{0.06\textwidth}p{0.08\textwidth}p{0.07\textwidth}p{0.07\textwidth}p{0.08\textwidth}p{0.08\textwidth}p{0.07\textwidth}p{0.08\textwidth}p{0.08\textwidth}p{0.02\textwidth}p{0.06\textwidth}}
\\
\hline
Range & $F_1$ & $\alpha_1$ & $\alpha_2$ & $t_{b_1}$ & $F_2$ & $\alpha_3$ & $\alpha_4$ & $t_{b_1}$ & $n$ & $\chi^2$ / d.f \\
& mJy &  &  & $\times10^{-3}$ day & mJy & & & day & & \\
\hline
X-ray & $0.52 \pm 0.10$ & $-1.2 \pm 1.0$ & $2.7 \pm 0.5$ & $1.9 \pm 0.4$ & $0.06 \pm 0.03$ & $0.5 \pm 0.2$ & $1.6 \pm 0.1$ & $0.02 \pm 0.01$ & 1* & 0.34 \\
Optical & $10 \pm 3$ & $-1.2*$ & $1.9 \pm 0.2$ & $1.9*$ & $0.27 \pm 0.04$ & $0.3 \pm 0.1$ & $1.64 \pm 0.08$ & $0.15 \pm 0.03$ & $1*$ & $1.1$ \\
\hline
\label{table:ag_fit}
\end{tabular}
\end{table*}

\subsection{Redshift estimation}

To estimate the redshift, we used the technique of simultaneously fitting the spectral energy distribution in the optical and X-ray bands under absorption conditions \citep{sha2010}. The time-sliced XRT X-ray spectrum and the UVOT optical images were
retrieved via the Swift Burst Analyser service. The spectra span the time interval 1554 -- 1638 s relative to $T_0$, where no significant color evolution is observed. The data were processed with the \texttt{HEASOFT}\footnote{\url{https://heasarc.gsfc.nasa.gov/ftools}} \citep{bla1995} software package. Figure \ref{fig:z_spec} presents a multiwavelength spectrum of the GRB 200829A afterglow fitted by the model curve (see below).

\begin{figure}
\centering
\includegraphics[width=0.45\textwidth]{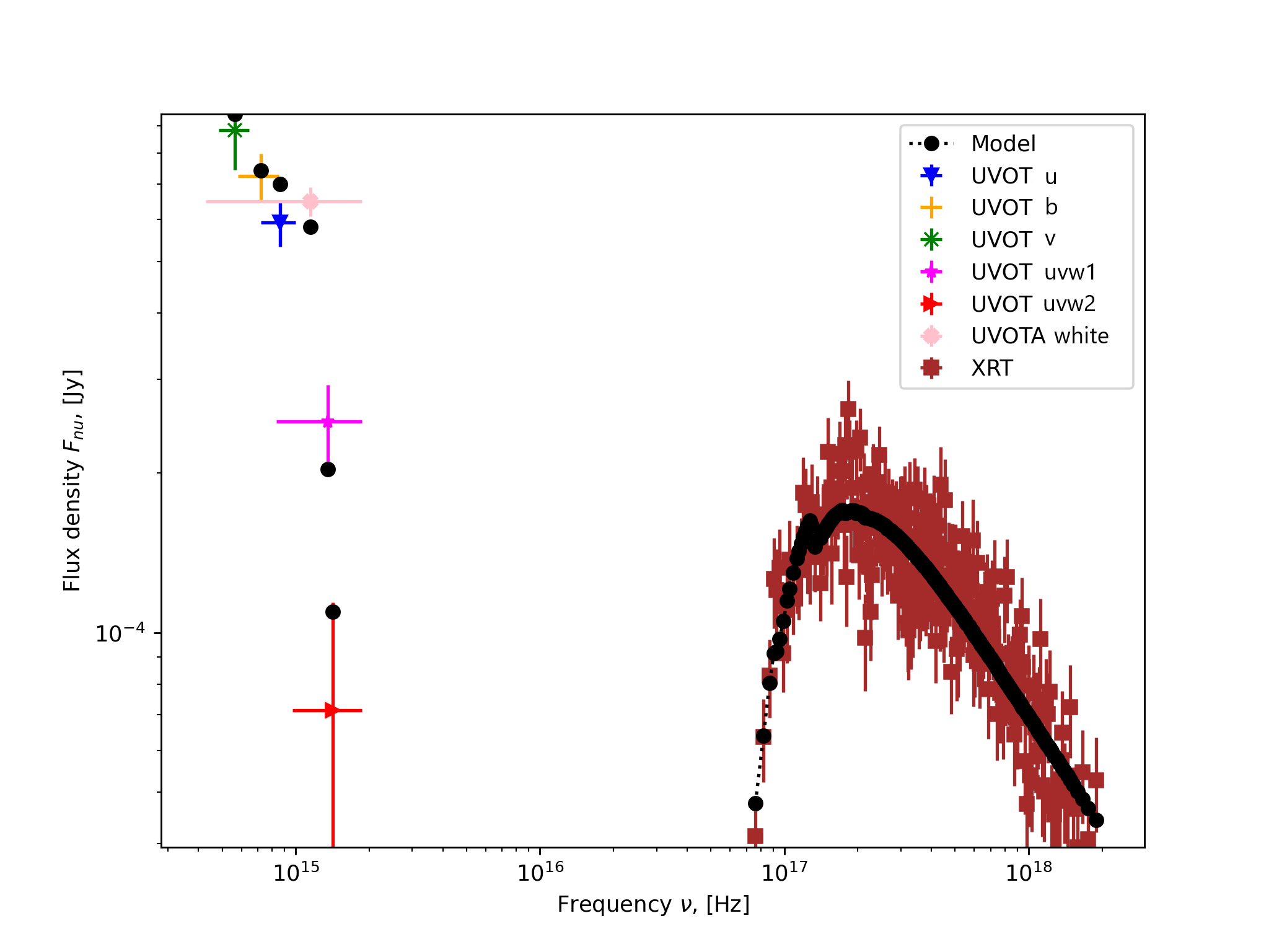}
\caption{Spectrum of GRB 200829A in the interval 1554 -- 1638 s relative to $T_0$. The emission frequency (Hz) is along the horizontal axis, while flux (Jy) is along the vertical axis. The model is designated by the black circles, while the remaining symbols mark the data.}\label{fig:z_spec}
\end{figure}

As the spectral model we use a power law and a broken power law. The model takes into account the optical extinction by interstellar dust grains and the photoelectric X-ray absorption and is represented by Eq. (\ref{eq:phot_z_spec_model}) in terms of \texttt{XSPEC} (\texttt{HEASOFT}), where the \texttt{zdust} components are responsible for the optical extinction. 

\begin{equation}\label{eq:phot_z_spec_model}
   \texttt{F} = \texttt{zdust} \times \texttt{zdust} \times \texttt{phabs} \times \texttt{zphabs} \times \texttt{powerlaw}
\end{equation}

These components are based on an empirical dust extinction law \citep{pei1992}. 
Three extinction laws are available in \texttt{XSPEC}: the Milky Way (MW), the Large Magellanic Cloud (LMC), and the Small Magellanic Cloud (SMC). 
The extinction model is characterized by the color excess $E(B - V)$, the total-to-selective extinction ratio $R(V) = A(V) / E(B - V)$, where $A(V)$ is the extinction, and the redshift $z$. 
The \texttt{phabs} and \texttt{zphabs} components allow the absorption by hydrogen in the Galaxy and the burst host galaxy to be taken into account. 
Both components are defined by the column density $N_H$ in units of $10^{-22}$ cm$^{-2}$, while \texttt{zphabs} is also defined by the redshift $z$. 
The \texttt{powerlaw} component is responsible for the power law spectrum and is specified by the spectral index $\beta$ and the normalization $K$ (photons$^{-1}$ keV$^{-1}$ cm${-2}$ at 1 keV). During our fitting the Galactic extinction parameters were fixed: $E(B - V) = 0.0364$ \citep{sch1998}, $R(V) = 3.08$ \citep{pei1992}, $A(V) = R(V)E(B - V) = 0.20$, and $N_H = 0.039 \times 10^{-22}$  cm$^{-2}$ \citep{hi4pi2016}. 
The redshift of the Galaxy was also fixed ($z = 0$). We found that the spectral data cannot be fitted with the broken power law and, therefore, this model was not considered. Table \ref{tab:phot_z_params} presents the results of fitting the spectral data for the early afterglow of GRB 200829A by the model defined in Eq. (\ref{eq:phot_z_spec_model}). The best spectral fit from the standpoint of the $\chi^2 / d.f.$ test is achieved when choosing the MW extinction law for the GRB host galaxy (see Table \ref{tab:phot_z_params}). The estimate of the photometric redshift $z = 1.29 \pm 0.04$ consistent with the independent estimate of $z = 1.25 \pm 0.02$ obtained previously by \cite{oat2020} by the same method corresponds to it. However, in
contrast to \cite{oat2020}, we also provide our estimates of the extinction parameters in the burst host galaxy. Note that in both cases the error is statistical. Below, $z = 1.29 \pm 0.04$ will be deemed to be the redshift of GRB 200829A.

\begin{table*}
\caption{Results of fitting the multiwavelength spectrum of the early afterglow. }
\centering
\begin{tabular}[H]{p{0.15\textwidth}p{0.1\textwidth}p{0.1\textwidth}p{0.1\textwidth}p{0.1\textwidth}p{0.1\textwidth}p{0.1\textwidth}p{0.1\textwidth}} %{lccccccc}
\\
\hline
Extinction law & $E(B-V)$ & $R(V)$ & $A(V)$ & $z$ & $nH$ & $\beta$ & $\chi^2 / d.f.$\\
 & mag &  & mag & & $10^{22}$ cm$^{-2}$ & &\\
\hline
MW & $0.36 \pm 0.02$ & 3.08* & $1.11 \pm 0.02$ & $1.29 \pm 0.04$ & $0.26 \pm 0.03$ & $1.72 \pm 0.02$ & $248 / 236$\\
LMC & $0.31 \pm 0.21$ & 3.16* & $0.98 \pm 0.21$ & $0.40 \pm 0.75$ & $0.42 \pm 0.59$ & $1.56 \pm 0.01$ & $329 / 236$\\
SMC & $0.19 \pm 0.04$ & 2.93* & $0.56 \pm 0.04$ & $1.36 \pm 0.33$ & $0.17 \pm 0.06$ & $1.60 \pm 0.02$ & $316 / 236$\\
\hline
\end{tabular}
\label{tab:phot_z_params}
\footnotesize{The values of $E(B-V)$, $R(V)$, $A(V)$, and $N_H$ are given for the host galaxy, and the corresponding values for the Galaxy were fixed: $E(B-V) = 0.0364$ mag, $R(V) = 3.08$, $A(V) = 0.20$ mag, and $N_H = 0.039 \times 10^{-22}$ cm$^{-2}$. The values of the fixed parameters are denoted with "*" symbol.}
\end{table*}

\section{Host galaxy}

\begin{figure}
\centering
\includegraphics[width=0.45\textwidth]{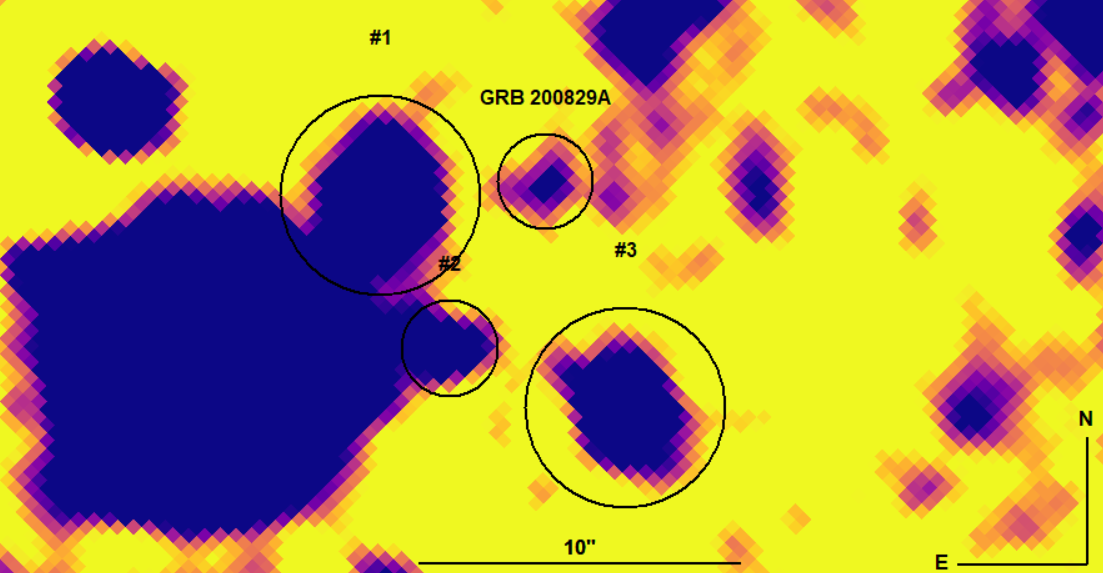}
\caption{Fragment of the $R$-band image of the GRB 200829A field obtained with the BTA telescope at SAO RAS. A diffuse object, the host galaxy, is seen at the location of the transient.}\label{fig:bta_host}
\end{figure}

We detected the host galaxy of GRB 200829A and calculated its contribution to the optical light curve using the observation with the Big Telescope Azimuthal (BTA) at SAO RAS $\sim 2$ years after the GRB. The BTA observation was carried out on July 31, 2022, at 18:30 UT with SCORPIO-1 (with a E2V CCD42-40 array) in the $R$ filter. A host galaxy candidate, whose coordinates coincided with those of GRB 200829A within the $\pm 0.1^{\prime\prime}$ error limits and whose brightness uncorrected for the Galactic extinction was $R = 25.5^{+0.4}_{-0.3}$, was found in the image with a total exposure time of 2910 s. Figure \ref{fig:bta_host} presents a fragment of the image in which the host galaxy candidate is marked.

The apparent magnitude after its correction for the Galactic extinction $A_R = 0.083$ \citep{schl2011} is $R = 25.4^{+0.4}_{
-0.3}$. This magnitude is consistent with the magnitude of a sample of host galaxies for $z = 1.29$ (see, e.g., \citealt{pozanenko_2008AstL}). The absolute magnitude of the host
galaxy is $M_R \sim -19.44$ for $z = 1.29$. Let us estimate the star formation rate (SFR) and the host mass $M_{host}$ from the relation $\log{SFR} = -0.381M_r - 8.029 \pm 0.486$ $M_{\odot}$ yr$^{-1}$, where $M_r$ is the absolute $r$ magnitude of a spiral galaxy (see, e.g., \citealt{mah2018}). The absolute magnitude of the GRB 200829A host galaxy will be $M_r = -19.20$, given the conversion of the magnitude from $R$ to $r$ ($r = R + 0.21$), and the correction for the Galactic extinction $A_r = 0.088$ \citep{schl2011}. The SFR estimate will then be $log{SFR} \approx -0.715 \pm 0.486 M_{\odot}$ yr$^{-1}$, whence we can estimate the host mass as $SFR \times T_{age}$ , where $T_{age}$ is
the host age. We obtain an estimate of the host mass
$M_{host} \sim 8.5 \times 10^9 M_{\odot}$. For comparison, the absolute magnitude of the GRB 181201A host galaxy is $M_R =
-18.5 \pm 0.2$, while its mass is $\sim 1.2 \times 10^9 M_{\odot}$ \citep{bel2020} (from the results of modelling the
galaxy spectrum). 

\begin{figure}
\centering
\includegraphics[width=0.45\textwidth]{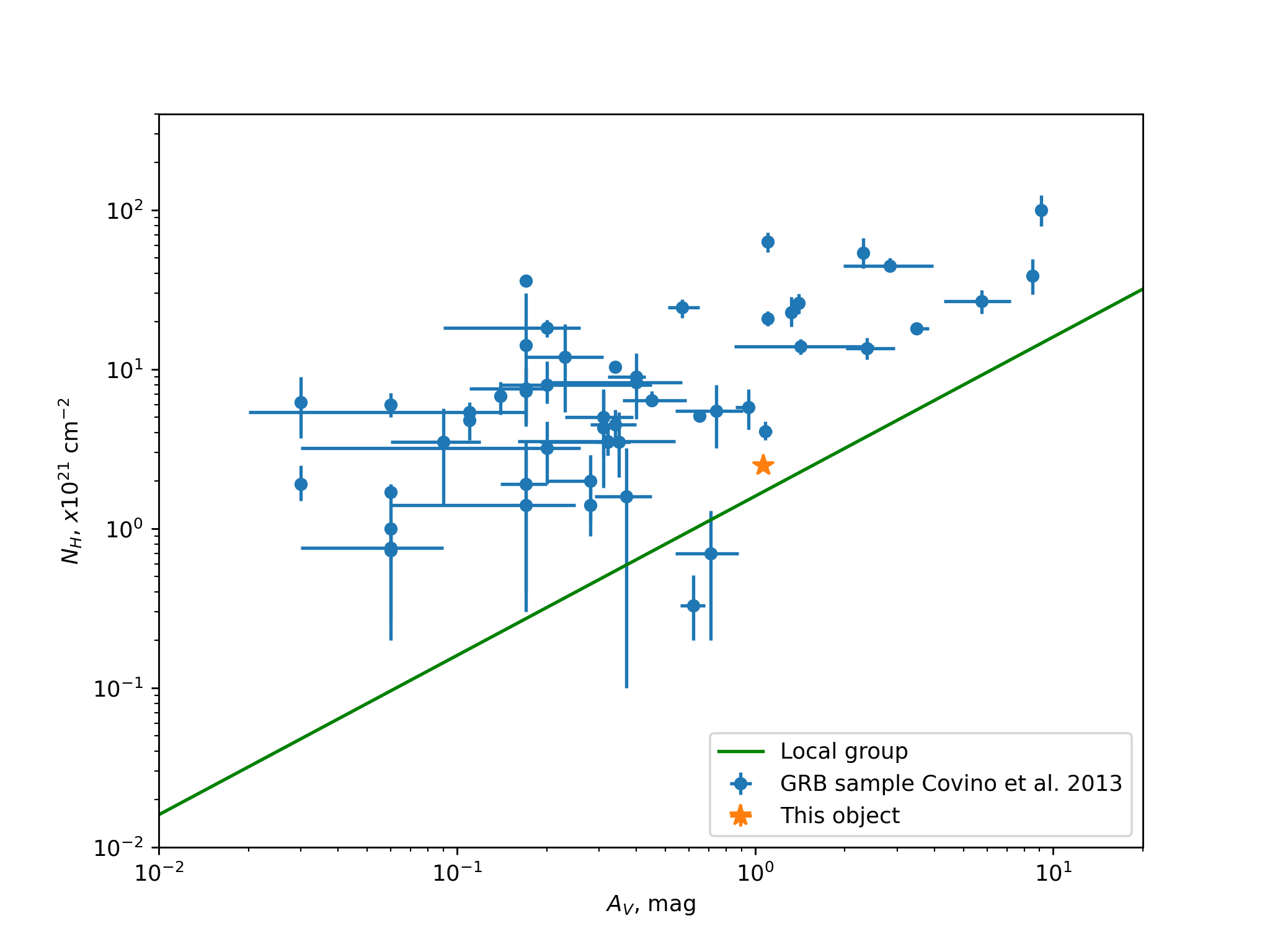}
\caption{Plot of the $N_H - A(V)$ relation. The blue squares mark the values from the GRB host galaxies \citep{cov2013}, while
the green line marks the linear law $\log(y) = 1.6 \log(x)$ that the overwhelming majority of galaxies in the sample match. The
red star designates $N_H -- A(V)$ for the GRB 200829A host galaxy.}\label{fig:nh_av}
\end{figure}

The previously obtained spectral fitting results (see Table \ref{tab:phot_z_params}) also allow one to determine the ratio
$N_H / A(V)$, from which the similarity of dust properties in various galaxies to be judged. If the dust properties are the same in the Universe, then the ratio $N_H / A(V)$ must increase linearly with distance to the galaxy. For example, for the host galaxy of GRB 200829A $N_H / A(V) \approx 2.4 \times 10^{22}$ cm$^{-2}$. The column density-to-extinction ratio for the host galaxy of GRB 200829A is consistent with most host galaxies of other GRBs represented in the sample \citep{cov2013}, for which $N_H / A(V) \approx 1.6 \times 10^{21}$ cm$^{-2}$, as is clearly shown in Fig. \ref{fig:nh_av}.

\section{Possible SN}

The long GRB 200829A could be accompanied by an SN. The SN search campaign was undertaken at the AZT-22 MAO telescope in the period 10 -- 46 days after the GRB detection. This period corresponds to a typical SN peak (8 -- 40 days) in the source frame (see, e.g. \citealt{bel2020}). During the observations the upper limit was $R = 24.0$ ($3\sigma$); however, no SN was detected. The upper limit for the absolute magnitude of a possible SN at $z = 1.29$ is $M_R > -23.0$. Here we made a correction for the extinction in the Galaxy $A_{R,MW} = 0.083$ and the GRB host galaxy $A_{\lambda} = 2.14 \pm 0.04$, where $\lambda =
0.641 \times (1 + z)$ $\mu$m is the effective wavelength of the
R filter at redshift $z$. $A_{\lambda}$ was calculated from the
following formula Eq.(\ref{eq:a_lambda}):

\begin{equation}\label{eq:a_lambda}
    A_V \simeq \frac{E(B-R)}{\xi (\frac{\lambda_{B}}{1+z}) \xi (\frac{\lambda_R}{1+z})}
\end{equation}

where, $E(B-R)$ -- is the color index between the $B$ and $R$ filters, $\xi$ -- is the extinction law, a $\lambda_{B}$ and  $\lambda_R$, effective wavelengths of the filters $B$, and $R$, respectively.

Our estimate is consistent with the absolute magnitude distribution of SNe from GRBs \citep{bel2020}.

\section{Discussion}

\subsection{Chromatic behavior of the light curve}
The multiwavelength light curve (Fig. \ref{fig:total_lc}) of GRB 200829A has an asynchronous (chromatic) behavior in the early afterglow that encompasses the irregularity (flare) and continues up until the jet break. The chromaticity during the flare can be interpreted in terms of the model of a structured jet (see, e.g.,
\citealt{duq2022}). According to this model, harder emission is observed closer to the jet axis and, for this reason, the X-ray emission corresponds to the part of the jet that is characterized by a larger gamma factor. Let us estimate the gamma factor of the jet from the following formula (see, e.g., \citealt{han2022}):

\begin{equation}\label{eq:gamma_factor}
    \Gamma_0 = 2\bigg[\frac{3E_{\gamma}(1+z)^3}{32 \pi n m_p c^5 \eta t_p^3 }\bigg]^{1/8}
\end{equation}

where, $m_p$ -- is the proton mass, $c$ -- is the speed of light, $E_{\gamma} = 2\pi (1 - \cos{\theta_j}) E_{iso}$ -- is the gamma-ray energy of the jet corrected for its opening angle $\theta_j$, $\eta = E_{\gamma} / \xi E_{iso}$, $t_p = t_b -\alpha_1 / \alpha_2)^{1/{\omega (\alpha_2 - \alpha_1)}}$, $t_b$ -- is the break time relative to  $T_0$ (days), $\alpha_1$, $\alpha_2$ -- are the power-law slopes of the light curve before and after the break, respectively, and $\omega = 1$. Note that in the light curve the jet break cannot be unambiguously determined.
Therefore, we will choose $t_b$ from Table \ref{table:ag_fit} in such a way that the open angle is not an extreme one. For example, 

\begin{enumerate}
    \item $t_b = T_0 + 1.9\times10^{-3}$ days corresponds to the flare peak in the early afterglow,
    \item $t_b = T_{0} + 0.02$ days corresponds to the break of the second component in the X-ray band,
    \item $t_b = T_{0} + 0.15$ days corresponds to the break of the second component in the optical band.
\end{enumerate}

The opening angle of the jet cone can be determined from the formula given below \citep[e.g.][]{sar1999,zha2007}:

\begin{equation}\label{eq:theta_jet_ism}
    \theta_j \sim 0.161{(\frac{t_b}{1+z})}^{3/8}{(\frac{\xi E_{52}}{n})}^{-1/8}
\end{equation}

where, $t_b$ is the break time relative to $T_0$ (days), $z$ is the redshift, $E_{52}$ is the isotropic energy $E_{iso}$ in units of $10^{52}$ erg, $n = 1$ cm$^{-3}$ is the ISM density, and the kinetic energy-to-radiation conversion factor is $\xi = 0.1$ \citep[e.g.][]{zha2007}. The jet opening angle will then be: $\theta_j(a) \approx 0.474^{+0.035}_{-0.039}$ $^{\circ}$, $\theta_j(b) \approx 1.15^{+0.47}_{-0.32} $ $^{\circ}$ or $\theta_j(c) \approx 2.44^{+0.71}_{-0.35}$ $^{\circ}$. The typical angle $\theta_j$determined for bursts with a jet break \citep[e.g.][]{wan2018} is known to be $\theta_j \approx 2.5 \pm 1.0 $ $^{\circ}$. Thus, the inhomogeneity (flare)
in the afterglow light curve is probably not connected
with the jet break since the opening angle $\theta_j(a) \sim 0.5 ^{\circ}$ would be anomalously narrow. In cases b) and c) he values turn out to be appropriate. At the same time, the opening angle derived from the X-ray data is narrower than from the optical ones. In cases (b) and (c) we will estimate the energy contained within the jet from the formula $E_{\gamma} = 2\pi (1 - \cos{\theta_j}) E_{iso}$. We will obtain, $E_{\gamma, O}  = 7.04^{+4.72}_{-1.85} \times 10^{51}$ erg when estimated from the optical light curve and $E_{\gamma, X} = 1.55^{+1.51}_{-0.75} \times 10^{51}$ erg when estimated from the X-ray
light curve. The gamma factors determined from the X-ray, $\Gamma_X = 197^{+170}_{-99}$, and optical, $\Gamma_O = 120^{+85}_{-34}$ light curves then formally coincide within the error limits.

\subsection{Inhomogeneity in the early afterglow}

\begin{figure}
\centering
\includegraphics[width=0.45\textwidth]{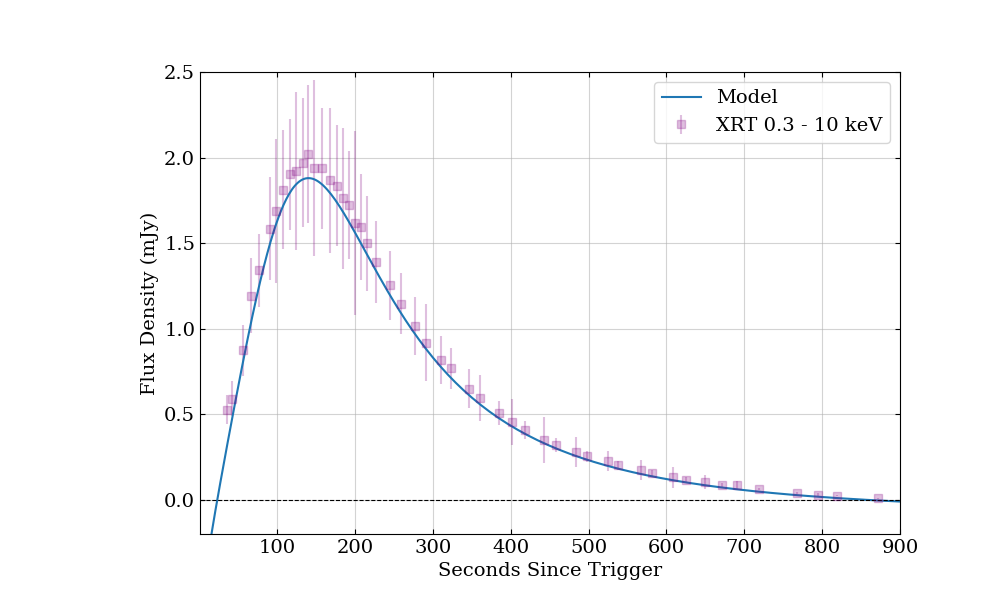}
\caption{Profile of the flare in the X-ray light curve of GRB 20089A after the afterglow model subtraction. The spectral flux
density in mJy is along the vertical axis, while the time since $T_0$ in seconds is along the horizontal axis. The model is designated by the blue solid line, while the observational data are indicated by the violet squares. The data corresponding to the powerful pulse in the burst prompt phase are not shown.}\label{fig:xray_flash}
\end{figure}

Our previous analysis showed that the light curve of the GRB 200829A afterglow contains a flare-type inhomogeneity at the early phase. The flares in GRB light curves were considered, for example, in the following papers \cite{pir2005,per2006,swe2013,yi2017,maz2018}. For instance, \cite{swe2013} investigated the duration as a function of the flare peak time $\Delta t / t_{peak}$ based on a sample of bursts from the second UVOT/Swift catalog. During the analysis it was established that $\Delta t / t_{peak} < 0.5$ for more than 80\% of the bursts from the sample. \cite{yi2017} and \cite{maz2018} also established that the flares have a linear $FWHM$ -- $t_{peak}$ correlation as well. Figure \ref{fig:xray_flash} presents the profile of the flare in the X-ray light curve of GRB 200829A obtained by subtracting the afterglow model.

The flare being investigated in this paper has $FWHM \approx 146$ s and $t_{peak} \approx 132$ s. The values obtained are consistent with the $FWHM$ -- $t_peak$ correlation \citep{yi2017,maz2018} and
are probably further evidence for the unified physical nature of the flares in GRB light curves. The flares may arise at the shock fed by the burst central engine \citep{bur2005b,pas2007,bar2021}, from the simultaneous observation of different structured-jet zones \citep{ben2020,duq2022}, or a variable accretion rate $\dot{M}$ \citep[e.g.][]{per2006}.

Our analysis of the first $\sim$ 250 s of the X-ray light curve for GRB 200829A (Fig. \ref{fig:bat}) does not reject the hypothesis about a plateau either. Owing to its strong magnetic field, the newly formed magnetar is believed to provide energy pumping and to maintain the luminosity at an approximately constant level for some time equal to the plateau duration \citep{pas2007,met2011,row2013}. Let us determine the energy release of the plateau in the X-ray light curve of GRB 200829A. First, we need to determine the plateau fluence in the 0.3–10 keV energy band ($\Delta \nu = 7.3 \times 10^{16}$ -- $2.4 \times 10^{18}$ Hz) from Eq. (\ref{eq:plateu_fluence})

\begin{equation}\label{eq:plateu_fluence}
    F_{plat} = \int_{t_1}^{t_2} f_{\nu}(t)dt \Delta \nu
\end{equation}

where the times $t_1 = 51.8$ s, and $t_2 = 86.4$ s specify
the light curve segment with a power-law slope $\sim 0$, corresponding to the plateau, and $f_{\nu}(t)$ is the spectral
flux density. Having substituted all of the necessary values, we obtain the plateau fluence $F_{plat} = 1.3 \times 10^{-10}$ erg cm$^{-2}$. The plateau energy in the 0.3 -- 10 keV band calculated from Eq. (\ref{eq:plateu_fluence}), will then be $E_{plateu} = 5.8 \times 10^{47}$ erg. Note that such values
are consistent with the $L_X$ -- $T_a$, where $L_x$ -- is the plateau luminosity in the 0.3 - 10 keV, and $T_a = t_2 / (1+z)$ is the time at the end of the plateau stage in the source frame \citep{dai2008,dai2021}.

\begin{table*}
    \centering
    \caption{Parameters of GRB 200829A}
    \begin{minipage}{1\textwidth}
    \begin{tabular}{lccccccccc}
        \hline
        R.A. & Dec & $z$ & $E_{iso}$ & $\theta_{j, O}$ & $\theta_{j, X}$ & $\Gamma_O$ & $\Gamma_X$ & $E_{\gamma, O}$ & $E_{\gamma, X}$\\
        Hours & Degrees & & $10^{54}$ erg & $^{\circ}$ & $^{\circ}$ & & & $10^{51}$ erg & $10^{51}$ erg\\
        \hline
        \\
        16:44:49.14 & +72:19:45.63 & $1.29 \pm 0.04$ & $1.30 \pm +0.01$ & $2.44^{+0.18}_{-0.20}$ & $1.13^{+0.20}_{-0.27}$ & $122^{+77}_{-23}$ & $199^{+157}_{-49}$ & $7.22^{+1.18}_{-1.20}$ & $1.59^{+0.60}_{-0.66}$\\
        \hline
    \end{tabular}
    \end{minipage}
    \label{tab:grb_properties}
\end{table*}

\begin{table*}
    \centering
    \caption{Parameters of the host galaxy}
    \begin{tabular}{lcccc}
        \hline
        $M_{R, Host}$ & $\log SFR$ & $M_{Host}$ & $A(V)$ & $N_{H}$\\
        mag & $M_{\odot}$ yr$^{-1}$ & $\times 10^9$ $M_{\odot}$ & mag & $\times 10^{21}$ cm$^{-2}$\\
        \hline
        $-19.44$ & $-0.715 \pm 0.486 $ & $\sim 8.5$ & $1.11 \pm 0.02$ & $2.6 \pm 0.3$\\
        \hline
    \end{tabular}
    \label{tab:host_properties}
\end{table*}

\section{Conclusions}

The parameters of GRB 200829A and its host galaxy determined in this paper are given in Tables \ref{tab:grb_properties} and \ref{tab:host_properties}, respectively.

Note some properties of GRB 200829A. In particular, the gamma-ray light curve exhibits a complex structure that is a superposition of pulses with different spectral hardness. Therefore, a proper
determination of the spectral lag turned out to be impossible. Our analysis of the gamma-ray BAT/Swift, GBM/Fermi, and SPI-ACS/INTEGRAL data showed that the event belongs to the class of long GRBs (Minaev and Pozanenko 2020a). GRB 200829A is among other $\sim$15 most powerful events in isotropic equivalent energy ($E_{iso} \gtrsim 10^{54}$ erg). A plateau gradually transitioning into extended emission with a duration $\sim$800 s is traced in the BAT/Swift light curve, while the flux decays
as a power law with a slope $\alpha \sim -1$ typical for an afterglow \citep{mozg21}.

Based on the results of UVOT/Swift and XRT/Swift observations, we made an independent photometric redshift estimate for GRB 200829A,
$z = 1.29 \pm 0.04$. 

No signature of SN from GRB 200829A was detected during its optical observations within 10 -- 40 days after $T_0$. The upper limit on the absolute magnitude of SN at maximum $M_R > -23$ is
consistent with the well-known absolute magnitude distribution of SNe associated with GRBs \citep{bel2022}.

The host galaxy of GRB 200829A was found with the BTA telescope at SAO RAS. Its absolute magnitude corrected for the Galactic extinction is $M_R \sim -19.44$ at $z = 1.29$, the star formation rate is $\log{SFR} \sim -0.7$ $M_{\odot}$ yr$^{-1}$, and the mass is $\sim 8 \times 10^{9}$ $M_{\odot}$. The absorption on the line of sight is consistent with the linear correlation $N_H \gtrsim 1.6 A(V)$ observed in GRB host galaxies \citep{cov2013}.

A joint analysis of the optical and X-ray observations of the GRB 200829A afterglow showed a chromatic behavior of the early afterglow before the break in the optical light curve (about 0.15 days since the GBM/Fermi trigger). The presence of a chromatic
feature is consistent with the model of a structured jet.

\section*{Acknowledgments}

N.S. Pankov, A.S. Pozanenko, P.Yu. Minaev, S.O. Belkin, E.V. Klunko, and A.A. Volnova thank the Ministry of Education and Science of the Russian Federation for its financial support, project no. 075-15-2022-1221 (2022-BRICS-8847-2335). This work made use of data supplied by the UK Swift Science Data Centre at the University of Leicester. Based on observations made with the Nordic Optical Telescope, owned in collaboration by the University of Turku and Aarhus University, and operated jointly by Aarhus University, the University of Turku and the University of Oslo, representing Denmark, Finland and Norway, the University of Iceland and Stockholm University at the Observatorio del Roque de los Muchachos, La Palma, Spain, of the Instituto de Astrofisica de Canarias.

\bibliographystyle{mnras}
\bibliography{biblio}

\appendix

\begin{table*}
\caption{Log of optical observations. The magnitudes in the UVOT/Swift $u$, $b$, $v$, $uvw1$, $uvw2$, $white$, $Clear$, $R$, and $I$ filters are given in the Vega photometric system, those in the $g$, $r$, $i$, and $z$ filters are given in the AB photometric system. The values are given without any correction for the extinction in the Galaxy and the GRB 200829A host galaxy. Continue in Table \ref{table:phot_data2}.}
\label{table:phot_data1}
\begin{tabular}[h]{cccccc}
\hline
$t-T_0$ & Magnitude & Flux density & Observatory/Telescope & Filter & GCN no. \\
days & & mJy & & & \\
\hline
0.002361 & $13.88 \pm 0.08$ & $10.213^{+0.777}_{-0.777}$ & Swift/UVOT & v & - \\
0.003391 & $14.33 \pm 0.07$ & $3.566^{+0.689}_{-0.689}$ & Swift/UVOT & white & - \\
0.003391 & $14.33 \pm 0.07$ & $3.566^{+0.689}_{-0.689}$ & Swift/UVOT & white & - \\
0.005810 & $14.55 \pm 0.05$ & $2.189^{+0.104}_{-0.104}$ & Swift/UVOT & u & - \\
0.007431 & $15.82 \pm 0.08$ & $1.906^{+0.134}_{-0.134}$ & Swift/UVOT & b & - \\
0.007720 & $15.46 \pm 0.08$ & $1.263^{+0.248}_{-0.248}$ & Swift/UVOT & white & - \\
0.008009 & $17.04 \pm 0.28$ & $0.113^{+0.029}_{-0.029}$ & Swift/UVOT & uvw2 & - \\
0.008299 & $15.43 \pm 0.10$ & $2.454^{+0.230}_{-0.230}$ & Swift/UVOT & v & - \\
0.008866 & $15.46 \pm 0.14$ & $0.582^{+0.074}_{-0.074}$ & Swift/UVOT & uvw1 & - \\
0.009155 & $15.08 \pm 0.09$ & $1.339^{+0.103}_{-0.103}$ & Swift/UVOT & u & - \\
0.009444 & $16.13 \pm 0.09$ & $1.430^{+0.112}_{-0.112}$ & Swift/UVOT & b & - \\
0.009734 & $15.66 \pm 0.08$ & $1.049^{+0.206}_{-0.206}$ & Swift/UVOT & white & - \\
0.010023 & $18.37 \pm 0.52$ & $0.033^{+0.016}_{-0.016}$ & Swift/UVOT & uvw2 & - \\
0.010301 & $15.89 \pm 0.12$ & $1.599^{+0.182}_{-0.182}$ & Swift/UVOT & v & - \\
0.010880 & $15.63 \pm 0.14$ & $0.497^{+0.067}_{-0.067}$ & Swift/UVOT & uvw1 & - \\
0.011944 & $16.00 \pm 0.07$ & $0.768^{+0.148}_{-0.148}$ & Swift/UVOT & white & - \\
0.013275 & $16.20 \pm 0.14$ & $1.205^{+0.157}_{-0.157}$ & Swift/UVOT & v & - \\
0.013843 & $16.08 \pm 0.17$ & $0.330^{+0.053}_{-0.053}$ & Swift/UVOT & uvw1 & - \\
0.014120 & $15.55 \pm 0.10$ & $0.869^{+0.077}_{-0.077}$ & Swift/UVOT & u & - \\
0.014410 & $16.39 \pm 0.10$ & $1.131^{+0.097}_{-0.097}$ & Swift/UVOT & b & - \\
0.014688 & $16.29 \pm 0.09$ & $0.590^{+0.117}_{-0.117}$ & Swift/UVOT & white & - \\
0.015324 & $16.32 \pm 0.20$ & $1.081^{+0.198}_{-0.198}$ & Swift/UVOT & v & - \\
0.015845 & $16.13 \pm 0.18$ & $0.316^{+0.051}_{-0.051}$ & Swift/UVOT & uvw1 & - \\
0.016123 & $15.80 \pm 0.11$ & $0.693^{+0.066}_{-0.066}$ & Swift/UVOT & u & - \\
0.016458 & $16.47 \pm 0.12$ & $1.049^{+0.117}_{-0.117}$ & Swift/UVOT & b & - \\
0.016701 & $16.26 \pm 0.09$ & $0.602^{+0.119}_{-0.119}$ & Swift/UVOT & white & - \\
0.016991 & $18.24 \pm 0.48$ & $0.037^{+0.017}_{-0.017}$ & Swift/UVOT & uvw2 & - \\
0.017280 & $16.41 \pm 0.16$ & $0.991^{+0.141}_{-0.141}$ & Swift/UVOT & v & - \\
0.017847 & $16.84 \pm 0.24$ & $0.163^{+0.036}_{-0.036}$ & Swift/UVOT & uvw1 & - \\
0.018148 & $16.03 \pm 0.11$ & $0.559^{+0.058}_{-0.058}$ & Swift/UVOT & u & - \\
0.018438 & $16.82 \pm 0.11$ & $0.759^{+0.077}_{-0.077}$ & Swift/UVOT & b & - \\
0.018727 & $16.42 \pm 0.09$ & $0.519^{+0.103}_{-0.103}$ & Swift/UVOT & white & - \\
0.019306 & $16.63 \pm 0.17$ & $0.809^{+0.128}_{-0.128}$ & Swift/UVOT & v & - \\
0.019873 & $16.36 \pm 0.19$ & $0.255^{+0.046}_{-0.046}$ & Swift/UVOT & uvw1 & - \\
0.020150 & $15.99 \pm 0.11$ & $0.582^{+0.060}_{-0.060}$ & Swift/UVOT & u & - \\
0.020440 & $16.81 \pm 0.11$ & $0.768^{+0.077}_{-0.077}$ & Swift/UVOT & b & - \\
0.035551 & $16.29 \pm 0.13$ & $0.864^{+0.100}_{-0.113}$ & TShAO/Zeiss-1000 & R & - \\
0.036296 & $16.34 \pm 0.08$ & $0.828^{+0.061}_{-0.066}$ & TShAO/Zeiss-1000 & R & - \\
0.037041 & $16.25 \pm 0.12$ & $0.898^{+0.092}_{-0.103}$ & TShAO/Zeiss-1000 & R & - \\
0.037785 & $16.37 \pm 0.11$ & $0.805^{+0.076}_{-0.084}$ & TShAO/Zeiss-1000 & R & - \\
0.038524 & $16.41 \pm 0.10$ & $0.773^{+0.066}_{-0.072}$ & TShAO/Zeiss-1000 & R & - \\
0.039271 & $16.36 \pm 0.08$ & $0.810^{+0.059}_{-0.064}$ & TShAO/Zeiss-1000 & R & - \\
0.040015 & $16.37 \pm 0.07$ & $0.802^{+0.053}_{-0.057}$ & TShAO/Zeiss-1000 & R & - \\
0.040756 & $16.43 \pm 0.07$ & $0.763^{+0.046}_{-0.049}$ & TShAO/Zeiss-1000 & R & - \\
0.041500 & $16.42 \pm 0.05$ & $0.771^{+0.037}_{-0.039}$ & TShAO/Zeiss-1000 & R & - \\
0.042240 & $16.45 \pm 0.05$ & $0.747^{+0.034}_{-0.035}$ & TShAO/Zeiss-1000 & R & - \\
0.042984 & $16.45 \pm 0.06$ & $0.750^{+0.041}_{-0.043}$ & TShAO/Zeiss-1000 & R & - \\
0.043734 & $16.49 \pm 0.08$ & $0.718^{+0.053}_{-0.057}$ & TShAO/Zeiss-1000 & R & - \\
0.044479 & $16.48 \pm 0.07$ & $0.724^{+0.046}_{-0.050}$ & TShAO/Zeiss-1000 & R & - \\
0.045224 & $16.42 \pm 0.07$ & $0.765^{+0.050}_{-0.054}$ & TShAO/Zeiss-1000 & R & - \\
0.045982 & $16.51 \pm 0.06$ & $0.704^{+0.035}_{-0.037}$ & TShAO/Zeiss-1000 & R & - \\
0.046728 & $16.52 \pm 0.07$ & $0.699^{+0.045}_{-0.049}$ & TShAO/Zeiss-1000 & R & - \\
0.047472 & $16.55 \pm 0.06$ & $0.682^{+0.038}_{-0.041}$ & TShAO/Zeiss-1000 & R & - \\
0.048218 & $16.54 \pm 0.08$ & $0.691^{+0.050}_{-0.054}$ & TShAO/Zeiss-1000 & R & - \\
0.048963 & $16.58 \pm 0.06$ & $0.662^{+0.038}_{-0.040}$ & TShAO/Zeiss-1000 & R & - \\
0.049708 & $16.54 \pm 0.07$ & $0.689^{+0.041}_{-0.044}$ & TShAO/Zeiss-1000 & R & - \\
0.050451 & $16.52 \pm 0.05$ & $0.697^{+0.032}_{-0.034}$ & TShAO/Zeiss-1000 & R & - \\
\hline
% \multicolumn{6}{p{16cm}}{References: GCN 28308 -- \cite{poz2020}, GCN 28316 -- \cite{vin2020}, GCN 28324 -- \cite{mos2020}, GCN 28324 -- \cite{zhu2020}, GCN 28328 -- \cite{mos2020b}, GCN 28330 -- \cite{zhu2020b}, GCN 28331 -- \cite{izz2020},  GCN 28333 -- \cite{vol2020}\\
\end{tabular}
\end{table*}

\begin{table*}
\caption{Log of optical observations. Continue in Table \ref{table:phot_data3}.}
\label{table:phot_data2}
\begin{tabular}[h]{cccccc}
\hline
$t-T_0$ & Magnitude & Flux density & Observatory/Telescope & Filter & GCN no. \\
days & & mJy & & & \\
\hline
0.051197 & $16.51 \pm 0.07$ & $0.706^{+0.043}_{-0.046}$ & TShAO/Zeiss-1000 & R & - \\
0.051941 & $16.55 \pm 0.08$ & $0.679^{+0.048}_{-0.052}$ & TShAO/Zeiss-1000 & R & - \\
0.052685 & $16.64 \pm 0.08$ & $0.628^{+0.046}_{-0.050}$ & TShAO/Zeiss-1000 & R & - \\
0.053429 & $16.66 \pm 0.07$ & $0.618^{+0.037}_{-0.040}$ & TShAO/Zeiss-1000 & R & - \\
0.054173 & $16.64 \pm 0.08$ & $0.630^{+0.043}_{-0.046}$ & TShAO/Zeiss-1000 & R & - \\
0.054916 & $16.66 \pm 0.08$ & $0.618^{+0.043}_{-0.047}$ & TShAO/Zeiss-1000 & R & - \\
0.055662 & $16.64 \pm 0.08$ & $0.624^{+0.046}_{-0.050}$ & TShAO/Zeiss-1000 & R & - \\
0.056408 & $16.69 \pm 0.07$ & $0.599^{+0.035}_{-0.037}$ & TShAO/Zeiss-1000 & R & - \\
0.057151 & $16.68 \pm 0.06$ & $0.602^{+0.034}_{-0.036}$ & TShAO/Zeiss-1000 & R & - \\
0.057897 & $16.67 \pm 0.07$ & $0.610^{+0.040}_{-0.042}$ & TShAO/Zeiss-1000 & R & - \\
0.058641 & $16.72 \pm 0.06$ & $0.583^{+0.029}_{-0.030}$ & TShAO/Zeiss-1000 & R & - \\
0.059387 & $16.66 \pm 0.08$ & $0.616^{+0.044}_{-0.048}$ & TShAO/Zeiss-1000 & R & - \\
0.060131 & $16.71 \pm 0.07$ & $0.588^{+0.034}_{-0.036}$ & TShAO/Zeiss-1000 & R & - \\
0.060873 & $16.74 \pm 0.08$ & $0.571^{+0.039}_{-0.042}$ & TShAO/Zeiss-1000 & R & - \\
0.061146 & $17.50 \pm 0.08$ & $0.192^{+0.038}_{-0.038}$ & Swift/UVOT & white & - \\
0.061617 & $16.75 \pm 0.08$ & $0.566^{+0.039}_{-0.042}$ & TShAO/Zeiss-1000 & R & - \\
0.062361 & $16.73 \pm 0.07$ & $0.575^{+0.035}_{-0.038}$ & TShAO/Zeiss-1000 & R & - \\
0.063109 & $16.75 \pm 0.08$ & $0.569^{+0.040}_{-0.043}$ & TShAO/Zeiss-1000 & R & - \\
0.065139 & $16.76 \pm 0.09$ & $0.560^{+0.046}_{-0.051}$ & TShAO/Zeiss-1000 & R & - \\
0.065887 & $16.77 \pm 0.07$ & $0.555^{+0.035}_{-0.037}$ & TShAO/Zeiss-1000 & R & - \\
0.066630 & $16.77 \pm 0.07$ & $0.554^{+0.036}_{-0.038}$ & TShAO/Zeiss-1000 & R & - \\
0.067374 & $16.82 \pm 0.06$ & $0.531^{+0.027}_{-0.028}$ & TShAO/Zeiss-1000 & R & - \\
0.068041 & $16.88 \pm 0.19$ & $0.502^{+0.079}_{-0.093}$ & Kitab-ISON/RC36 & Clear & - \\
0.068117 & $16.92 \pm 0.08$ & $0.483^{+0.033}_{-0.035}$ & TShAO/Zeiss-1000 & R & - \\
0.068863 & $16.78 \pm 0.07$ & $0.551^{+0.034}_{-0.036}$ & TShAO/Zeiss-1000 & R & - \\
0.069605 & $16.77 \pm 0.09$ & $0.555^{+0.043}_{-0.046}$ & TShAO/Zeiss-1000 & R & - \\
0.070350 & $16.86 \pm 0.07$ & $0.515^{+0.030}_{-0.032}$ & TShAO/Zeiss-1000 & R & - \\
0.071093 & $16.86 \pm 0.08$ & $0.512^{+0.035}_{-0.037}$ & TShAO/Zeiss-1000 & R & - \\
0.071839 & $16.81 \pm 0.05$ & $0.536^{+0.024}_{-0.025}$ & TShAO/Zeiss-1000 & R & - \\
0.072582 & $16.89 \pm 0.09$ & $0.499^{+0.041}_{-0.045}$ & TShAO/Zeiss-1000 & R & - \\
0.073053 & $16.79 \pm 0.21$ & $0.548^{+0.098}_{-0.120}$ & Kitab-ISON/RC36 & Clear & - \\
0.073328 & $16.89 \pm 0.09$ & $0.498^{+0.039}_{-0.043}$ & TShAO/Zeiss-1000 & R & - \\
0.074074 & $16.92 \pm 0.08$ & $0.485^{+0.036}_{-0.039}$ & TShAO/Zeiss-1000 & R & - \\
0.074818 & $16.87 \pm 0.08$ & $0.508^{+0.036}_{-0.039}$ & TShAO/Zeiss-1000 & R & - \\
0.075561 & $16.88 \pm 0.09$ & $0.503^{+0.038}_{-0.041}$ & TShAO/Zeiss-1000 & R & - \\
0.076305 & $16.95 \pm 0.09$ & $0.473^{+0.038}_{-0.041}$ & TShAO/Zeiss-1000 & R & - \\
0.077049 & $16.90 \pm 0.09$ & $0.493^{+0.041}_{-0.044}$ & TShAO/Zeiss-1000 & R & - \\
0.077791 & $16.96 \pm 0.09$ & $0.465^{+0.039}_{-0.042}$ & TShAO/Zeiss-1000 & R & - \\
0.078076 & $17.11 \pm 0.18$ & $0.407^{+0.061}_{-0.072}$ & Kitab-ISON/RC36 & Clear & - \\
0.078532 & $16.97 \pm 0.07$ & $0.463^{+0.031}_{-0.033}$ & TShAO/Zeiss-1000 & R & - \\
0.079277 & $16.94 \pm 0.07$ & $0.478^{+0.030}_{-0.032}$ & TShAO/Zeiss-1000 & R & - \\
0.080020 & $16.93 \pm 0.07$ & $0.481^{+0.032}_{-0.034}$ & TShAO/Zeiss-1000 & R & - \\
0.080766 & $17.03 \pm 0.09$ & $0.438^{+0.033}_{-0.036}$ & TShAO/Zeiss-1000 & R & - \\
0.081412 & $17.65 \pm 0.08$ & $0.168^{+0.033}_{-0.033}$ & Swift/UVOT & white & - \\
0.081512 & $16.95 \pm 0.07$ & $0.470^{+0.031}_{-0.034}$ & TShAO/Zeiss-1000 & R & - \\
0.082256 & $17.05 \pm 0.08$ & $0.428^{+0.032}_{-0.034}$ & TShAO/Zeiss-1000 & R & - \\
0.083000 & $17.01 \pm 0.08$ & $0.446^{+0.033}_{-0.035}$ & TShAO/Zeiss-1000 & R & - \\
0.083099 & $17.27 \pm 0.28$ & $0.350^{+0.079}_{-0.103}$ & Kitab-ISON/RC36 & Clear & - \\
0.083745 & $17.02 \pm 0.09$ & $0.441^{+0.035}_{-0.038}$ & TShAO/Zeiss-1000 & R & - \\
0.084490 & $17.00 \pm 0.10$ & $0.452^{+0.040}_{-0.044}$ & TShAO/Zeiss-1000 & R & - \\
0.085234 & $17.06 \pm 0.09$ & $0.425^{+0.033}_{-0.036}$ & TShAO/Zeiss-1000 & R & - \\
0.085977 & $17.03 \pm 0.09$ & $0.438^{+0.034}_{-0.037}$ & TShAO/Zeiss-1000 & R & - \\
0.086720 & $16.99 \pm 0.09$ & $0.456^{+0.038}_{-0.041}$ & TShAO/Zeiss-1000 & R & - \\
0.087465 & $17.00 \pm 0.06$ & $0.449^{+0.024}_{-0.026}$ & TShAO/Zeiss-1000 & R & - \\
0.088110 & $17.31 \pm 0.29$ & $0.340^{+0.079}_{-0.102}$ & Kitab-ISON/RC36 & Clear & - \\
0.088206 & $17.04 \pm 0.07$ & $0.433^{+0.025}_{-0.027}$ & TShAO/Zeiss-1000 & R & - \\
0.088952 & $16.97 \pm 0.06$ & $0.463^{+0.025}_{-0.026}$ & TShAO/Zeiss-1000 & R & - \\
0.089692 & $17.08 \pm 0.09$ & $0.418^{+0.034}_{-0.037}$ & TShAO/Zeiss-1000 & R & - \\
0.090437 & $17.09 \pm 0.08$ & $0.413^{+0.028}_{-0.030}$ & TShAO/Zeiss-1000 & R & - \\
\hline
% \multicolumn{6}{p{16cm}}{References: GCN 28308 -- \cite{poz2020}, GCN 28316 -- \cite{vin2020}, GCN 28324 -- \cite{mos2020}, GCN 28324 -- \cite{zhu2020}, GCN 28328 -- \cite{mos2020b}, GCN 28330 -- \cite{zhu2020b}, GCN 28331 -- \cite{izz2020},  GCN 28333 -- \cite{vol2020}\\
\end{tabular}
\end{table*}

\begin{table*}
\caption{Log of optical observations. Continue in Table \ref{table:phot_data4}.}
\label{table:phot_data3}
\begin{tabular}[h]{cccccc}
\hline
$t-T_0$ & Magnitude & Flux density & Observatory/Telescope & Filter & GCN no. \\
days & & mJy & & & \\
\hline
0.091182 & $17.05 \pm 0.09$ & $0.431^{+0.033}_{-0.036}$ & TShAO/Zeiss-1000 & R & - \\
0.091927 & $17.09 \pm 0.07$ & $0.413^{+0.027}_{-0.029}$ & TShAO/Zeiss-1000 & R & - \\
0.092672 & $17.04 \pm 0.09$ & $0.435^{+0.036}_{-0.039}$ & TShAO/Zeiss-1000 & R & - \\
0.093134 & $16.81 \pm 0.17$ & $0.535^{+0.076}_{-0.088}$ & Kitab-ISON/RC36 & Clear & - \\
0.093926 & $17.14 \pm 0.08$ & $0.397^{+0.029}_{-0.031}$ & TShAO/Zeiss-1000 & R & - \\
0.094673 & $17.19 \pm 0.10$ & $0.379^{+0.032}_{-0.036}$ & TShAO/Zeiss-1000 & R & - \\
0.095420 & $17.19 \pm 0.08$ & $0.377^{+0.028}_{-0.030}$ & TShAO/Zeiss-1000 & R & - \\
0.096163 & $17.00 \pm 0.06$ & $0.452^{+0.025}_{-0.027}$ & TShAO/Zeiss-1000 & R & - \\
0.098157 & $16.86 \pm 0.17$ & $0.510^{+0.074}_{-0.087}$ & Kitab-ISON/RC36 & Clear & - \\
0.103180 & $16.82 \pm 0.22$ & $0.532^{+0.098}_{-0.121}$ & Kitab-ISON/RC36 & Clear & - \\
0.108203 & $17.15 \pm 0.22$ & $0.391^{+0.072}_{-0.088}$ & Kitab-ISON/RC36 & Clear & - \\
0.113215 & $17.11 \pm 0.23$ & $0.407^{+0.079}_{-0.097}$ & Kitab-ISON/RC36 & Clear & - \\
0.118238 & $17.39 \pm 0.27$ & $0.314^{+0.068}_{-0.087}$ & Kitab-ISON/RC36 & Clear & - \\
0.119339 & $17.29 \pm 0.08$ & $0.344^{+0.024}_{-0.026}$ & TShAO/Zeiss-1000 & R & - \\
0.120315 & $17.33 \pm 0.07$ & $0.333^{+0.022}_{-0.023}$ & TShAO/Zeiss-1000 & R & - \\
0.121292 & $17.26 \pm 0.08$ & $0.356^{+0.025}_{-0.026}$ & TShAO/Zeiss-1000 & R & - \\
0.122265 & $17.39 \pm 0.06$ & $0.314^{+0.018}_{-0.019}$ & TShAO/Zeiss-1000 & R & - \\
0.123244 & $17.33 \pm 0.07$ & $0.334^{+0.020}_{-0.022}$ & TShAO/Zeiss-1000 & R & - \\
0.123261 & $17.17 \pm 0.24$ & $0.385^{+0.077}_{-0.097}$ & Kitab-ISON/RC36 & Clear & - \\
0.124219 & $17.43 \pm 0.08$ & $0.302^{+0.021}_{-0.022}$ & TShAO/Zeiss-1000 & R & - \\
0.125192 & $17.30 \pm 0.08$ & $0.341^{+0.026}_{-0.028}$ & TShAO/Zeiss-1000 & R & - \\
0.126167 & $17.44 \pm 0.10$ & $0.301^{+0.026}_{-0.029}$ & TShAO/Zeiss-1000 & R & - \\
0.127143 & $17.39 \pm 0.08$ & $0.314^{+0.021}_{-0.023}$ & TShAO/Zeiss-1000 & R & - \\
0.128123 & $17.45 \pm 0.08$ & $0.299^{+0.022}_{-0.024}$ & TShAO/Zeiss-1000 & R & - \\
0.128284 & $17.16 \pm 0.22$ & $0.389^{+0.071}_{-0.087}$ & Kitab-ISON/RC36 & Clear & - \\
0.129104 & $17.37 \pm 0.07$ & $0.321^{+0.019}_{-0.021}$ & TShAO/Zeiss-1000 & R & - \\
0.130078 & $17.38 \pm 0.09$ & $0.318^{+0.025}_{-0.027}$ & TShAO/Zeiss-1000 & R & - \\
0.131051 & $17.46 \pm 0.10$ & $0.294^{+0.026}_{-0.028}$ & TShAO/Zeiss-1000 & R & - \\
0.132032 & $17.45 \pm 0.10$ & $0.298^{+0.026}_{-0.029}$ & TShAO/Zeiss-1000 & R & - \\
0.133003 & $17.49 \pm 0.09$ & $0.287^{+0.023}_{-0.025}$ & TShAO/Zeiss-1000 & R & - \\
0.133978 & $17.34 \pm 0.08$ & $0.328^{+0.025}_{-0.027}$ & TShAO/Zeiss-1000 & R & - \\
0.134954 & $17.36 \pm 0.10$ & $0.322^{+0.027}_{-0.030}$ & TShAO/Zeiss-1000 & R & - \\
0.135935 & $17.40 \pm 0.07$ & $0.312^{+0.019}_{-0.020}$ & TShAO/Zeiss-1000 & R & - \\
0.136912 & $17.23 \pm 0.09$ & $0.364^{+0.029}_{-0.031}$ & TShAO/Zeiss-1000 & R & - \\
0.137886 & $17.36 \pm 0.09$ & $0.323^{+0.025}_{-0.028}$ & TShAO/Zeiss-1000 & R & - \\
0.138562 & $17.31 \pm 0.26$ & $0.340^{+0.073}_{-0.093}$ & Kitab-ISON/RC36 & Clear & - \\
0.141863 & $17.58 \pm 0.09$ & $0.224^{+0.018}_{-0.019}$ & Koshka (INASAN)/Zeiss-1000 & I & - \\
0.143123 & $17.72 \pm 0.09$ & $0.250^{+0.035}_{-0.036}$ & Koshka (INASAN)/Zeiss-1000 & R & - \\
0.147173 & $17.30 \pm 0.19$ & $0.342^{+0.054}_{-0.064}$ & Kitab-ISON/RC36 & Clear & - \\
0.148628 & $17.45 \pm 0.10$ & $0.296^{+0.027}_{-0.029}$ & TShAO/Zeiss-1000 & R & - \\
0.149606 & $17.44 \pm 0.14$ & $0.301^{+0.036}_{-0.040}$ & TShAO/Zeiss-1000 & R & - \\
0.150584 & $17.57 \pm 0.12$ & $0.266^{+0.028}_{-0.031}$ & TShAO/Zeiss-1000 & R & - \\
0.151559 & $17.36 \pm 0.11$ & $0.324^{+0.030}_{-0.033}$ & TShAO/Zeiss-1000 & R & - \\
0.152531 & $17.53 \pm 0.10$ & $0.276^{+0.024}_{-0.026}$ & TShAO/Zeiss-1000 & R & - \\
0.153510 & $17.52 \pm 0.11$ & $0.279^{+0.027}_{-0.029}$ & TShAO/Zeiss-1000 & R & - \\
0.154643 & $17.72 \pm 0.10$ & $0.249^{+0.036}_{-0.037}$ & Koshka (INASAN)/Zeiss-1000 & R & - \\
0.154643 & $17.71 \pm 0.13$ & $0.199^{+0.022}_{-0.025}$ & Koshka (INASAN)/Zeiss-1000 & I & - \\
0.158932 & $17.48 \pm 0.22$ & $0.289^{+0.053}_{-0.066}$ & Kitab-ISON/RC36 & Clear & - \\
0.169696 & $17.11 \pm 0.17$ & $0.408^{+0.060}_{-0.071}$ & Kitab-ISON/RC36 & Clear & - \\
0.176090 & $17.65 \pm 0.09$ & $0.210^{+0.017}_{-0.018}$ & SAO RAS/Zeiss-1000 & I & 28322 \\
0.177833 & $18.03 \pm 0.11$ & $0.211^{+0.047}_{-0.048}$ & CrAO/AZT-11 & R & - \\
0.180448 & $17.57 \pm 0.24$ & $0.266^{+0.053}_{-0.066}$ & Kitab-ISON/RC36 & Clear & - \\
0.182092 & $17.71 \pm 0.10$ & $0.233^{+0.021}_{-0.023}$ & CrAO/AZT-11 & R & - \\
0.182093 & $17.91 \pm 0.09$ & $0.231^{+0.047}_{-0.048}$ & CrAO/AZT-11 & R & - \\
0.185933 & $17.90 \pm 0.09$ & $0.167^{+0.013}_{-0.014}$ & Koshka (INASAN)/Zeiss-1000 & I & - \\
0.186352 & $17.81 \pm 0.10$ & $0.213^{+0.018}_{-0.020}$ & CrAO/AZT-11 & R & - \\
0.186353 & $18.01 \pm 0.12$ & $0.214^{+0.048}_{-0.049}$ & CrAO/AZT-11 & R & - \\
0.190413 & $18.00 \pm 0.08$ & $0.197^{+0.032}_{-0.033}$ & Koshka (INASAN)/Zeiss-1000 & R & - \\
0.190611 & $17.73 \pm 0.10$ & $0.229^{+0.020}_{-0.022}$ & CrAO/AZT-11 & R & - \\
\hline
% \multicolumn{6}{p{16cm}}{References: GCN 28308 -- \cite{poz2020}, GCN 28316 -- \cite{vin2020}, GCN 28324 -- \cite{mos2020}, GCN 28324 -- \cite{zhu2020}, GCN 28328 -- \cite{mos2020b}, GCN 28330 -- \cite{zhu2020b}, GCN 28331 -- \cite{izz2020},  GCN 28333 -- \cite{vol2020}\\
\end{tabular}
\end{table*}

\begin{table*}
\caption{Log of optical observations, last table.}
\label{table:phot_data4}
\begin{tabular}[h]{cccccc}
\hline
$t-T_0$ & Magnitude & Flux density & Observatory/Telescope & Filter & GCN no. \\
days & & mJy & & & \\
\hline
0.190613 & $17.99 \pm 0.11$ & $0.218^{+0.048}_{-0.049}$ & CrAO/AZT-11 & R & - \\
0.194870 & $17.80 \pm 0.10$ & $0.215^{+0.019}_{-0.021}$ & CrAO/AZT-11 & R & - \\
0.194873 & $18.11 \pm 0.13$ & $0.199^{+0.048}_{-0.049}$ & CrAO/AZT-11 & R & - \\
0.196467 & $17.68 \pm 0.21$ & $0.241^{+0.042}_{-0.051}$ & Kitab-ISON/RC36 & Clear & - \\
0.215673 & $18.04 \pm 0.10$ & $0.147^{+0.013}_{-0.014}$ & Koshka (INASAN)/Zeiss-1000 & I & - \\
0.220163 & $18.16 \pm 0.07$ & $0.171^{+0.031}_{-0.031}$ & Koshka (INASAN)/Zeiss-1000 & R & - \\
0.230043 & $18.24 \pm 0.09$ & $0.184^{+0.015}_{-0.016}$ & Konkoly/RC80 & r & - \\
0.230043 & $18.49 \pm 0.11$ & $0.146^{+0.014}_{-0.016}$ & Konkoly/RC80 & i & - \\
0.245413 & $18.27 \pm 0.09$ & $0.133^{+0.012}_{-0.013}$ & Koshka (INASAN)/Zeiss-1000 & I & - \\
0.249893 & $18.24 \pm 0.07$ & $0.162^{+0.031}_{-0.031}$ & Koshka (INASAN)/Zeiss-1000 & R & - \\
0.250093 & $18.56 \pm 0.04$ & $0.137^{+0.005}_{-0.005}$ & Nanshan/NEXT & r & 28324 \\
0.267623 & $18.50 \pm 0.05$ & $0.145^{+0.007}_{-0.007}$ & Nanshan/NEXT & i & 28324 \\
0.272891 & $18.91 \pm 0.05$ & $0.099^{+0.004}_{-0.005}$ & Liverpool & g & 28331 \\
0.275143 & $18.31 \pm 0.08$ & $0.115^{+0.008}_{-0.009}$ & Koshka (INASAN)/Zeiss-1000 & I & - \\
0.277891 & $18.55 \pm 0.04$ & $0.138^{+0.005}_{-0.005}$ & Liverpool & r & 28331 \\
0.279623 & $18.64 \pm 0.06$ & $0.150^{+0.031}_{-0.031}$ & Koshka (INASAN)/Zeiss-1000 & R & - \\
0.280923 & $18.89 \pm 0.04$ & $0.101^{+0.004}_{-0.004}$ & Nanshan/NEXT & g & 28324 \\
0.282891 & $18.65 \pm 0.04$ & $0.126^{+0.005}_{-0.005}$ & Liverpool & i & 28331 \\
0.287891 & $18.57 \pm 0.04$ & $0.136^{+0.005}_{-0.005}$ & Liverpool & z & 28331 \\
0.304883 & $18.51 \pm 0.11$ & $0.095^{+0.009}_{-0.010}$ & Koshka (INASAN)/Zeiss-1000 & I & - \\
0.309373 & $18.44 \pm 0.09$ & $0.137^{+0.031}_{-0.031}$ & Koshka (INASAN)/Zeiss-1000 & R & - \\
0.334613 & $18.53 \pm 0.09$ & $0.094^{+0.007}_{-0.008}$ & Koshka (INASAN)/Zeiss-1000 & I & - \\
0.339103 & $18.46 \pm 0.09$ & $0.135^{+0.031}_{-0.031}$ & Koshka (INASAN)/Zeiss-1000 & R & - \\
0.375793 & $18.85 \pm 0.12$ & $0.099^{+0.031}_{-0.031}$ & Koshka (INASAN)/Zeiss-1000 & R & - \\
0.400093 & $18.73 \pm 0.08$ & $0.078^{+0.006}_{-0.006}$ & Koshka (INASAN)/Zeiss-1000 & I & - \\
0.452193 & $18.90 \pm 0.08$ & $0.096^{+0.030}_{-0.030}$ & Koshka (INASAN)/Zeiss-1000 & R & - \\
0.602836 & $20.06 \pm 0.12$ & $0.018^{+0.004}_{-0.004}$ & Swift/UVOT & white & - \\
1.062923 & $>20.4$ & $>0.02535$ & Nanshan/NEXT & r & 28330 \\
1.150123 & $20.37 \pm 0.13$ & $0.020^{+0.002}_{-0.003}$ & SAO RAS/Zeiss-1000 & R & 28322 \\
1.167123 & $20.14 \pm 0.10$ & $0.021^{+0.002}_{-0.002}$ & SAO RAS/Zeiss-1000 & I & 28322 \\
1.291593 & $20.14 \pm 0.20$ & $0.021^{+0.004}_{-0.004}$ & Koshka (INASAN)/Zeiss-1000 & I & 28333 \\
1.293093 & $20.99 \pm 0.25$ & $0.029^{+0.028}_{-0.028}$ & Koshka (INASAN)/Zeiss-1000 & R & 28333 \\
1.307923 & $20.99 \pm 0.16$ & $0.015^{+0.002}_{-0.002}$ & Nanshan/NEXT & r & 28330 \\
1.346603 & $21.01 \pm 0.05$ & $0.014^{+0.001}_{-0.001}$ & NOT & r & - \\
1.361891 & $21.17 \pm 0.09$ & $0.012^{+0.001}_{-0.001}$ & Liverpool & g & 28331 \\
1.365891 & $21.13 \pm 0.10$ & $0.013^{+0.001}_{-0.001}$ & Liverpool & r & 28331 \\
1.370891 & $21.25 \pm 0.09$ & $0.011^{+0.001}_{-0.001}$ & Liverpool & i & 28331 \\
1.375891 & $20.50 \pm 0.08$ & $0.023^{+0.002}_{-0.002}$ & Liverpool & z & 28331 \\
2.194383 & $21.54 \pm 0.40$ & $0.007^{+0.002}_{-0.003}$ & AbAO/AS-32 & R & - \\
2.728877 & $22.40 \pm 0.34$ & $0.002^{+0.001}_{-0.001}$ & Swift/UVOT & white & - \\
2.802903 & $22.40 \pm 0.08$ & $0.004^{+0.0005}_{-0.0005}$ & NOT & r & - \\
3.15717 & >20.4 & >0.021 & Mondy/AZT-33IK & R & -\\
5.20315 & >21.5  & >0.008 & Mondy/AZT-33IK & R & -\\
13.070211 & $>23.6$ & $>0.00103$ & MAO/AZT-22 & R & - \\
17.077277 & $>23.7$ & $>0.00094$ & MAO/AZT-22 & R & - \\
44.561496 & $>23.9$ & $>0.00078$ & MAO/AZT-22 & R & - \\
349.653193 & $>24.5$ & $>0.00057$ & MAO/AZT-22 & R & - \\
701.205657 & $25.46 \pm 0.35$ & $0.00020^{+0.00015}_{-0.00028}$ & SAO RAS/BTA & R & - \\
\hline
\multicolumn{6}{l}{References: GCN 28308 -- \cite{poz2020}, GCN 28316 -- \cite{vin2020}, GCN 28324 -- \cite{mos2020},}\\
\multicolumn{6}{l}{GCN 28324 -- \cite{zhu2020}, GCN 28328 -- \cite{mos2020b}, GCN 28330 -- \cite{zhu2020b}, GCN 28331 -- \cite{izz2020},}\\
\multicolumn{6}{l}{GCN 28333 -- \cite{vol2020}}\\
% \multicolumn{6}{p16cm}{}\\
\end{tabular}
\end{table*}

% Don't change these lines
% \bsp	% typesetting comment
\label{lastpage}
\end{document}